%% file: manuscript_single_vfinal.tex
\documentclass[12pt,journal,draftcls,a4paper,onecolumn]{IEEEtran} 
\usepackage{dsfont}

\usepackage{subeqnarray}
\usepackage{cleveref}
\usepackage{amssymb}
\usepackage{amsfonts}
\usepackage{amsmath}
\usepackage[dvipdfmx]{graphicx}
\usepackage{bmpsize}
\usepackage{amsmath}
\usepackage[all,poly]{xy}
\usepackage{multirow}
\usepackage{algorithm}
\usepackage{algorithmic}
\usepackage{color}
\usepackage[noadjust]{cite}

\title{Robust Spectral Unmixing of Sparse Multispectral Lidar Waveforms using Gamma Markov Random Fields}
\author{Yoann Altmann, Aurora Maccarone, Aongus McCarthy, Gregory Newstadt, Gerald S. Buller, Steve
McLaughlin and Alfred Hero
\thanks{Yoann Altmann, Aurora Maccarone, Aongus McCarthy, Gerald S. Buller and Steve
McLaughlin are with School of Engineering and Physical Sciences, Heriot-Watt University,
U.K. (email: \{Y.Altmann;am827;A.McCarthy; G.S.Buller;S.McLaughlin\}@hw.ac.uk).}\thanks{Gregory Newstadt is with Google Inc., Pittsburgh, PA, U.S.A. (email: newstage37@gmail.com).}\thanks{Alfred Hero is with Department of Electrical Engineering and Computer Science, University of Michigan, U.S.A. (email: hero@eecs.umich.edu).}
\thanks{This study was supported by EPSRC via grants EP/J015180/1, EP/K015338/1 and EP/M01326X/1 and by US DOE Consortium for Verification Technology DE-NA0002534 and US Army grant W911NF-14-1-0479.
}}

\newcommand{\blambda}{\boldsymbol{\lambda}}
\newcommand{\bbeta}{\boldsymbol{\beta}}

\newcommand{\btheta}{\boldsymbol{\theta}}
\newcommand{\bLambda}{\boldsymbol{\Lambda}}
\newcommand{\bGam}{\boldsymbol{\Gamma}}

\newcommand{\bphi}{{\boldsymbol \phi}}
\newcommand{\bPhi}{{\boldsymbol \Phi}}

\graphicspath{{figures/}}

\input notations_yoann.tex

\begin{document}
\maketitle

\begin{abstract}
This paper presents a new Bayesian spectral unmixing algorithm to analyse remote scenes sensed via sparse multispectral Lidar measurements. To a first approximation, in the presence of a target, each Lidar waveform consists of a main peak, whose position depends on the target distance and whose amplitude depends on the wavelength of the laser source considered (i.e, on the target reflectivity). Besides, these temporal responses are usually assumed to be corrupted by Poisson noise in the low photon count regime. When considering multiple wavelengths, it becomes possible to use spectral information in order to identify and quantify the main materials in the scene, in addition to estimation of the Lidar-based range profiles. Due to its anomaly detection capability, the proposed hierarchical Bayesian model, coupled with an efficient Markov chain Monte Carlo algorithm, allows robust estimation of depth images together with abundance and outlier maps associated with the observed 3D scene. The proposed methodology is illustrated via experiments conducted with real multispectral Lidar data acquired in a controlled environment. The results demonstrate the possibility to unmix spectral responses constructed from extremely sparse photon counts (less than 10 photons per pixel and band). 
\end{abstract}


\section{Introduction}
\label{sec:intro}
Laser altimetry (or Lidar) is an acknowledged tool for extracting spatial structure from three-dimensional (3D) scenes. Using time-of-flight to create a distance profile, signal analysis can recover, for instance, tree and canopy heights, leaf area indices and ground slope by analyzing the reflected photons from a target.
Conversely, passive multispectral (MSI) and hyperspectral images (HSI) are widely used to extract spectral information about the scene which can provide useful parameters about the composition and health condition of the canopy. The most natural method to extract spatial and spectral information from sensed scenes is to couple Lidar data and multi/hyperspectral images \cite{Dalponte2008,Buckley2013}. Although the fusion of Lidar data and HSIs can improve scene characterization, data synchronization issues in space (alignment, resolution) and time (dynamic scene, change of observation conditions, etc.) make this problematic and these are still open issues. For these reasons, multispectral Lidar (MSL) has recently received attention from the remote sensing community for its ability to extract both spatial and spectral information from 3D scenes. The key advantage of MSL is the ability to potentially provide information on the full 3D distribution of materials, especially for scenes including semi-transparent objects (e.g., vegetation or fences). When the Lidar return signal is sufficiently strong the received light field will exhibit easily separable spatial and spectral peaks corresponding to the different surfaces and material properties. In this case, classical methods for 3D reconstruction can be applied \cite{Wallace2012, Hakala2012, Wallace2014,Altmann2015a}. For instance, in  \cite{Hakala2012}, for each waveform, a series
of peak positions and reflectivity parameters is estimated iteratively by identifying and
subtracting sequentially the peak with the highest amplitude, until a pre-defined threshold
is reached. The pixels and spectral bands are processed independently, leading to one point
cloud per spectral band. In \cite{Wallace2012,Wallace2014}, a Bayesian approach is adopted to first estimate the number and positions of the peaks (from a single band) and these parameters are
then used to estimate the reflectivity parameters associated with the remaining spectral
bands. Another motivation for MSL is that HSIs, even when fully synchronized, can only integrate the spectral response along the path of each optical ray, and cannot measure the spectral response as a function of distance, e.g. depth into a forest canopy.

In \cite{Ramirez2012,Wallace2014,Altmann2015a}, spectral unmixing techniques were developed to analyze 3D scenes composed of multi-layered objects, assuming that the spectral signatures of the materials composing the scenes were known and assuming linear mixing processes. In contrast to \cite{Ramirez2012,Wallace2014}, the unmixing method proposed in \cite{Altmann2015a}, which consists of a Metropolis-within-Gibbs sampler, assumes that there is either a single peak (whose position is estimated) or multiple peaks whose positions are known. This method has been extended in \cite{Altmann2016whispers} to account for and identify possible deviations from the classical linear mixing model (LMM) used to estimate the amount/abundances of each endmember (assumed known) present in the scene. In this work, we further improve the robust unmixing method of \cite{Altmann2016whispers} in order to enhance abundance and range profile estimation when there are extremely low photon counts per pixel, which is relevant to situations where the acquisition time is very restricted, e.g., in extreme low light imaging. Specifically, we propose a new abundance prior model which promotes smooth abundance maps and we propose a depth prior model that promotes piece-wise homogeneous depth profiles. As will be shown in the experimental section of this paper, by accounting for the intrinsic spatial organization of natural images, our algorithm provides significantly better ranging and unmixing results when compared to pixel-wise estimation algorithms, in particular when considering Lidar with extremely low photon counts.
 
Single-photon Lidar MSL systems usually record, for each pixel/region of the scene, a histogram of time delays between emitted laser pulses and the detected photon
arrivals. Due to the low number of photon arrivals detected, Poisson noise models are more appropriate for single-photon MSL data than Gaussian noise models that are typically used for HSIs in the high photon density regime. Such models are particularly relevant for challenging scenarios where the recorded waveforms consists of very few photons, i.e., less than $10$ on average across the image pixels for each wavelength, which occurs when reducing the overall acquisition time. In this paper, we demonstrate the ability of robust Bayesian Poisson unmixing methods to simultaneously estimate endmember fractions, extract depth information, and detect anomalous regions that are poorly represented by the assumed nominal LMM. The performance is illustrated on an experimental testbed with clay objects of different colours and an MSL imaging system ($33$ wavelengths ranging from $500$nm to $820$nm), under favourable observation conditions. 

The benefit of Bayesian approaches is that prior distributions can be chosen for the unknown parameters of the model, effectively smoothing the unmixing solution through regularization. For example, a total-variation (TV) regularization is proposed in \cite{Rudin1992,Chambolle2004} to regularize the depth estimation problem and gamma Markov random fields (MRFs) are used to model the spatial dependencies affecting the unknown abundance maps. As illustrated in \cite{Altmann2015a}, such Markovian models can be used to promote local spatial smoothness of the estimated abundances while providing enough flexibility to handle sharp transitions. Gamma-MRFs have the nice property of leading to conjugate prior models under Poisson noise assumption.

To allow for moderate deviations from the linear mixing model we propose a sparse anomaly model within a hierarchical Bayesian framework. Such anomalies can occur in the presence of scarcely represented materials or when there are local variations in the main spectral signatures of the scene. To capture these anomalies, we use a 3D Ising model for sparse deviations from the standard LMM. Although the proposed method is able to detect deviations from the LMM that are not necessarily sparse (such as endmember variability or mis-specification and nonlinear mixtures), such deviations are likely to yield locally biased abundance estimates. Nonetheless, the method that we propose can potentially be used to identify pixels for which more complex mixing models might be used.

The proposed prior models are incorporated into a hierarchical Bayesian model and the joint posterior distribution of these parameters is derived using Bayes' theorem. A simulation-based method is then developed to estimate the unknown parameters. More precisely, we construct a stochastic gradient
Markov chain Monte Carlo (SG-MCMC) algorithm to jointly generate samples according to the posterior of interest and adjust the MRF hyperparameters during the burn-in period of the sampler.  This strategy
has several important advantages in the context of estimating MRF hyperparameters whose conditional distributions are highly complex and have no closed-form expressions. Firstly, it allows for the automatic adjustment of the hyperparameters for each dataset and, secondly it has a computational cost that is several times lower than that of competing approaches, such as those that include the hyperparameters in the
Bayesian model through hierarchical priors. The proposed Bayesian approach produces a predictor of optimal estimator performance (through the derivation of posterior measures of uncertainty) while reducing potential convergence issues arising from the non-concavity of the log-posterior (due to the presence of anomaly terms). More sophisticated optimization techniques, e.g., variational methods, are worth additional study but are outwith the scope of this current paper.

The remainder of the paper is organized as follows. Section \ref{sec:model} introduces the observation model associated with MSL returns for a single-layered object to be analyzed. Section \ref{sec:bayesian} presents the hierarchical Bayesian model associated with the robust spectral unmixing problem considered and the associated posterior distribution. Section \ref{sec:MCMC} describes the SG-MCMC method used to sample from the posterior of interest and subsequently approximate appropriate Bayesian estimators. The relation between the proposed model and the Poisson Factor Analysis is addressed in Section \ref{sec:PFA}. Results of experiments conducted on real MSL data are shown and discussed in Section \ref{sec:simulations} and conclusions are reported in Section \ref{sec:conclusion}.

\vspace{-0.2cm}
\section{Problem formulation}
\label{sec:model} 
This section introduces the statistical model associated with MSL returns for a single-surface reflecting object which will be used in Section \ref{sec:bayesian} for robust spectral unmixing of MSL data. We consider a 4-D array $\MATpix$ of Lidar waveforms of dimension $N_{\textrm{row}} \times N_{\textrm{col}} \times L \times T$, where $N_{\textrm{row}}$ and $N_{\textrm{col}}$ stands for the number of rows and columns of the regular spatial sampling grid (in the transverse plane), $L$ is the number of spectral bands or wavelengths used to reconstruct the scene and $T$ is the number of temporal (corresponding to range) bins. Let $\Vpix{i,j,\ell} = \left[\MATpix \right]_{i,j,\ell,t} = [\pix{i,j,\ell}{1},\ldots,\pix{i,j,\ell}{\nbbin}]\transp$ be the Lidar waveform obtained in the pixel $(i,j)$ using the $\ell$th wavelength. The element $\pix{i,j,\ell}{\nobin}$ is the photon count within the $\nobin$th bin of the $\ell$th spectral band considered. Due to the design of the proposed experiments (performed indoors in this work), and to simplify the estimation problem, we further assume that ambient noise counts (e.g., from additional illumination sources or dark counts from the detector) can be neglected. Thus, for each pixel, the detected photons (originally emitted by the laser sources) only result from direct path reflections onto the surface of the object of interest. Moreover, we assume that the laser beam (for each pixel) encounters a single surface which is assumed to be locally orthogonal to the beam direction. This is typically the case for short to mid-range (up to dozens of meters) depth imaging where the divergence of the laser source(s) can be neglected. Let $d_{i,j}$ be the position of an object surface at a given range from the sensor, whose mean spectral signature (observed at $L$ wavelengths) is denoted as $\blambda_{i,j}=[\lambda_{i,j,1},\ldots,\lambda_{i,j,L}]\transp$.  
According to \cite{Hernandez2007,Altmann2016}, each photon count $\pix{i,j,\ell}{t}$ is assumed to be drawn from the following Poisson distribution 
\begin{eqnarray}
\label{eq:model0}
\pix{i,j,\ell}{\nobin}|\lambda_{i,j,\ell}, t_{i,j} \sim \mathcal{P}\left(\lambda_{i,j,\ell}g_{i,j,\ell}(\nobin-t_{i,j}) \right)
\end{eqnarray}
where $g_{i,j,\ell}(\cdot)$ is the photon impulse response, evaluated at discrete time positions as discussed in Section \ref{sec:priors} and whose shape can differ between wavelength channels and pixel locations. In Eq. \eqref{eq:model0}, $t_{i,j}$ is the characteristic time-of-flight of photons emitted by a pulsed laser source and reaching the detector after being reflected by a target at range $d_{i,j}$ ($d_{i,j}$ and $t_{i,j}$ are linearly related in free-space propagation). Moreover, the impulse responses $\left\lbrace g_{i,j,\ell}(\cdot)\right\rbrace$ are assumed to be known, as occurs when they can be accurately estimated during imaging system calibration. We further assume that the spectral signatures of the scene surfaces can be decomposed as linear mixtures of $R$ known spectral signatures $\Vmat{r}$ (also referred to as endmembers and gathered in the $L \times R$ matrix $\MATmat=[\Vmat{1},\ldots,\Vmat{R}]$) possibly corrupted by sparse anomalies (or deviations from the linear mixture), that is 
\begin{eqnarray}
\blambda_{i,j} = \MATmat\Vabond{i,j} + \bfr_{i,j}, \quad \forall i, j,
\end{eqnarray}
where $\Vabond{i,j}=[\abond{i,j}{1},\ldots,\abond{i,j}{R}]\transp$ contains the abundances of the $R$ endmembers in the pixel $(i,j)$ and $\bfr_{i,j} \in \mathbb{R}_+^L$ is a sparse vector that captures anomalies that do not fit the LMM $\blambda_{i,j} = \MATmat\Vabond{i,j}$. As explained in \cite{Newstadt2014ssp,Altmann2015c}, these anomalies in the resulting robust LMM can be due to actual outliers/corrupted data, nonlinear spectral mixtures or intrinsic endmember variability over the spectral bands. Note that due to physical considerations the unknown abundance vectors $\left\lbrace \Vabond{i,j}\right\rbrace_{i,j}$ can be assumed to have positive entries. It is important to recall that, in this work, we consider applications where the observed objects consist of a single visible surface per pixel. We do not consider cases where the photons can penetrate through objects (e.g., semi-transparent materials for which we would like to infer the internal composition) or be reflected from multiple surfaces. This assumption allows the spectral unmixing problem to be reduced to two spatial dimensions, which will be extended to distributed targets in future work. Moreover, we consider a single spectral signature per material, which does not depend on the object orientation with respect to the imaging system. Should the material signatures change significantly (due to the surface orientation or its intrinsic spectral variability), the changes are expected to be captured by the vectors $\bfr_{i,j}$.
The problem addressed is to jointly estimate the range of the targets (for all the image pixels) and to solve the robust spectral unmixing problem (e.g., estimating the abundance vectors and identifying the pixels corrupted by anomalies). The next section introduces the Bayesian model for this problem.

\vspace{-0.2cm}
\section{Bayesian model}
\label{sec:bayesian}
\subsection{Likelihood}
\label{sec:lik}
Assuming that the MSL waveforms $\Vpix{i,j} = \left\lbrace \pix{i,j,\ell}{\nobin} \right\rbrace_{\ell,t}$ associated with a given pixel $(i,j)$ result from photon reflection from a single surface associated with the spectrum $\blambda_{i,j}$ and according to \eqref{eq:model0}, the likelihood associated with the pixel $(i,j)$ can be expressed as
\begin{eqnarray}
\label{eq:sing_lik}
f(\Vpix{i,j}|\blambda_{i,j},t_{i,j}) =\prod_{\ell,t}f_{\mathcal{P}}(\pix{i,j,\ell}{\nobin};\lambda_{i,j,\ell}g_{i,j,\ell}(\nobin-t_{i,j})),
\end{eqnarray}
when it is assumed that the detected photon counts/noise realizations,
conditioned on their mean in all channels/spectral bands, are conditionally independent. Note that in \eqref{eq:sing_lik}, $f_{\mathcal{P}}(\cdot;\lambda)$ denotes the probability mass function of the Poisson distribution with mean $\lambda$. Considering that the noise realizations in the different pixels are also conditionally independent, the joint likelihood can be expressed as 
\begin{eqnarray}
\label{eq:joint_lik}
f(\MATpix|\bLambda,\bfT)=\prod_{i,j}f(\Vpix{i,j}|\blambda_{i,j},t_{i,j}),
\end{eqnarray}
where $\bLambda=\left\lbrace\blambda_{i,j}\right\rbrace_{i,j}$ and $\bfT$ is a matrix gathering the target ranges.
\subsection{Prior distributions}
\label{sec:priors}
\subsubsection{Range parameters}
Each target position is a discrete variable defined on $\bbT=\{t_{min},\ldots,t_{max}\}$, such that $1\leq t_{min}\leq t_{max} \leq T$. In this paper we set $(t_{min},t_{max})=(301,T-300)$ and the temporal resolution of the grid is set to the resolution of the single-photon detection (i.e., $2$ picoseconds (ps) in Section \ref{sec:simulations}). As in \cite{Altmann2016}, to  account  for  the  spatial  correlations  between  the neighbouring  pixels,  we  propose  to  use  a Markov random field  as  a  prior distribution for $t_{i,j}$ given its
neighbours $\bfT_{\mathcal{V}(i,j)}$, i.e., $f(t_{i,j}|\bfT_{\backslash (i,j)})=f(t_{i,j}|\bfT_{\mathcal{V}(i,j)})$ where $\mathcal{V}(i,j)$ is  the depth neighbourhood  of  the  pixel $(i,j)$, $\bfT_{\mathcal{V}(i,j)}=\{t_{i',j'}\}_{(i',j') \in \mathcal{V}(i,j)}$
and $\bfT_{\backslash (i,j)}=\{t_{i',j'}\}_{(i',j')\neq (i,j)}$. More  precisely, we propose the
following discrete MRF
\begin{eqnarray}
\label{eq:prior_t1}
f(\bfT|\epsilon) =\dfrac{1}{G(\epsilon)}\exp \left[-\epsilon \phi(\bfT) \right]
\end{eqnarray}
where $\epsilon\geq 0$ is a parameter tuning the amount of correlation
between pixels, $G(\epsilon)$ is a normalization (or partition) constant
and  where $\phi(\cdot)$ is  an  arbitrary  cost  function  modeling
correlation  between  neighbours.  In  this  work  we  propose  to
use the following cost function
\begin{eqnarray}
\label{eq:TV_reg}
\phi(\bfT) = \sum_{i,j} \sum_{(i',j')\in \mathcal{V}(i,j)} |t_{i,j}-t_{i',j'}|,
\end{eqnarray}
which   corresponds   to   a total-variation regularization  \cite{Rudin1992,Chambolle2004} promoting  piecewise  constant  depth  image.
Moreover,  the  higher  the  value  of $\epsilon$, the  more  correlated
the  neighboring  pixels.  Several  neighborhood  structures  can
be  employed  to  define $\mathcal{V}(i,j)$; here, a four pixel structure (1-order  neighbourhood)  will  be  considered  in  the  rest  of
the paper for the depth parameters. 

To illustrate the effect of this spatial regularization of the depth profile, we also consider an alternative depth prior model constructed from independent uniform priors
\begin{eqnarray}
\label{eq:prior_t2}
p(t_{i,j}=t) = \dfrac{1}{T'}, \quad t \in \bbT,
\end{eqnarray}
where $T'=\textrm{card}(\bbT)$. 

\subsubsection{Material abundances}
It follows from \eqref{eq:sing_lik} that gamma distributions are conjugate priors for the unknown abundances $\abond{i,j}{\nomat}$ (the resulting conditional distributions are mixtures of gamma distributions, see \cite{Altmann2016} for details). Consequently, we propose such gamma priors for these parameters. Note that such priors also ensure the positivity of the proposed estimator of abundances. In a similar manner to \cite{Altmann2016}, we assign $\abond{i,j}{\nomat}$ the following gamma prior
\begin{eqnarray}
\label{eq:prior_a}
\abond{i,j}{\nomat} \sim \mathcal{G}\left(c_r,\dfrac{\bar{a}_{i,j,r}}{c_r} \right)
\end{eqnarray} 
where $\bar{a}_{i,j,r}>0$ is a local parameter related to the prior mean of $\abond{i,j}{\nomat}$ and $c_r>0$ is a global parameter (i.e, one per endmember) which controls the shape of the distribution tails and thus the prior deviation of $\abond{i,j}{\nomat}$ from $\bar{a}_{i,j,r}$. Hierarchical Bayesian models generally allow the construction of elaborate prior models in which parameters can be related through the introduction of additional parameters which generally belong to higher levels in the Bayesian hierarchical model. For instance, setting $\bar{a}_{i,j,r}=\bar{a}_{r}$ in \eqref{eq:prior_a} reduces to choosing the same prior, characterized by $(c_r,\bar{a}_r)$, for all abundances associated with the $r$th endmember. Conversely, here we specify \eqref{eq:prior_a} to reflect the prior belief that abundances exhibit
spatial correlations. In particular, due to the spatial organization of images, we expect the
values of $\abond{i,j}{r}$ to vary smoothly from one pixel to another. In order to model this behaviour, 
we specify $\bar{a}_{i,j,r}$ such that the resulting prior for the abundances associated with each endmember $\MATabond_r=\left\lbrace \abond{i,j}{\nomat}\right\rbrace_{i,j}$ is a hidden gamma-MRF \cite{Dikmen2010}. 

\begin{figure}[ht]
\centering
\includegraphics[width=\columnwidth]{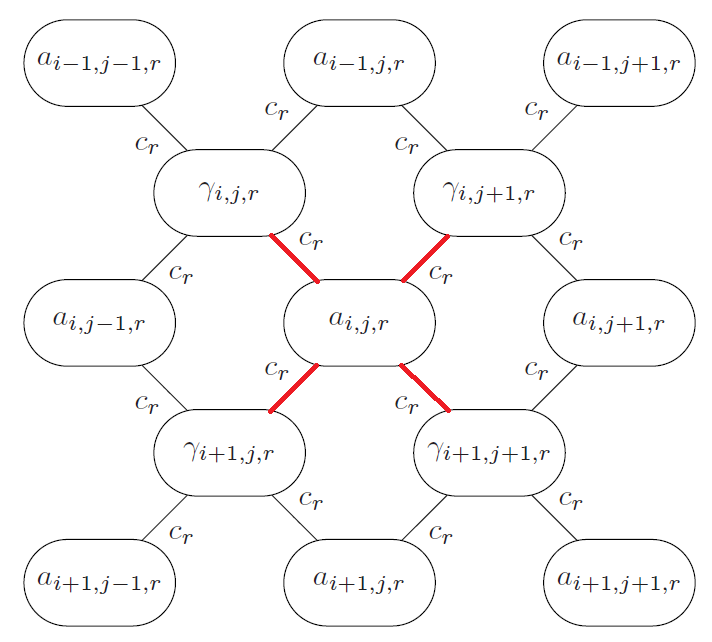}
\caption{Proposed $1$st order gamma-MRF neighborhood structure for the abundances $\{a_{i,j,r} \}, \forall (i,j) \in \mathcal{V}_{\MATabond_r}$ and $\forall r \in \{1,\ldots, R\}$. We set $a_{i,j,r}=0.01, \forall (i,j) \notin \mathcal{V}_{\MATabond_r}$}
\label{fig:neighbour_GMRF}
\end{figure}

More precisely, we introduce $R$ auxiliary matrices $\bGam_r$ of size $(N_{\textrm{row}}+1) \times (N_{\textrm{col}}+1)$, with elements $\gamma_{i,j,r} \in \mathbb{R}^+$ and define bipartite conditional independence graphs between $\MATabond_r$ and $\bGam_r$ such that each $\abond{i,j}{\nomat}$ is connected to four neighbour elements of $\bGam_r$ and vice-versa. This $1$st order neighbourhood structure is depicted in Fig. \ref{fig:neighbour_GMRF}, where we point out that any given $\abond{i,j}{\nomat}$ and $\abond{i+1,j}{\nomat}$ are $2$nd order neighbours via $\gamma_{i+1,j,r}$ and $\gamma_{i+1,j+1,r}$. Following the general gamma-MRF model proposed in \cite{Dikmen2010}, and specified here by the neighbourhood structure depicted in Fig. \ref{fig:neighbour_GMRF}, we assign a gamma-MRF prior for each $(\MATabond_r,\bGam_r)$, and obtain the following joint priors for $(\MATabond_r,\bGam_r)$
\begin{eqnarray}
\label{eq:GMRF}
f(\MATabond_r,\bGam_r|c_r) & = & \dfrac{1}{G(c_r)} \prod_{(i,j) 
\in \mathcal{V}_{\MATabond_r}} \abond{i,j}{r}^{\left(c_r-1 \right)}\nonumber\\
& \times & \prod_{(i',j') \in \mathcal{V}_{\bGam_r}} 
\left(\gamma_{i',j',r}\right)^{-\left(c_r+1 \right)} \nonumber\\
 & \times & \prod_{\left((i,j),(i',j')\right) \in \mathcal{E}} \exp 
\left(\dfrac{-c_r \abond{i,j}{r}}{4 \gamma_{i',j',r}} \right),
\end{eqnarray}
where $\mathcal{V}_{\MATabond_r}=\left\lbrace 1,\ldots,N_{\textrm{row}}\right \rbrace \times \left\lbrace 1,\ldots,N_{\textrm{col}}\right \rbrace$, $\mathcal{V}_{\bGam_r}=\left\lbrace 1,\ldots,N_{\textrm{row}}+1 \right \rbrace \times \left\lbrace 1,\ldots,N_{\textrm{col}}+1 \right \rbrace$, and the edge set $\mathcal{E}$ consists of pairs $\left((i,j),(i',j')\right)$ representing the connection between $\abond{i,j}{r}$ and $\gamma_{i',j',r}$.
It can be seen from \eqref{eq:GMRF} that 
\begin{subeqnarray}
\label{eq:prior1_r}
\slabel{eq:prior1_r2}
\abond{i,j}{r}|\bGam_r,c_r &\sim & \mathcal{G}\left(c_r, \dfrac{\bar{a}_{i,j,r}(\bGam_r)}{c_r}\right) \\
\slabel{eq:prior1_r3}
\gamma_{i,j,r}|\MATabond_r,c_r &\sim & \mathcal{IG}\left(c_r,c_r \beta_{i,j,r}(\bfR)\right)
\end{subeqnarray}
where
\begin{eqnarray*}
\bar{a}_{i,j,r}(\bGam) & = & 4\left(\gamma_{i,j,r}^{-1} + \gamma_{i-1,j,r}^{-1} + \gamma_{i,j-1,r}^{-1} + \gamma_{i-1,j-1,r}^{-1}\right)^{-1}\\
\beta_{i,j,r}(\bfR) &=& \left(\abond{i,j}{r} + \abond{i+1,j}{r} + \abond{i,j+1}{r} \abond{i+1,j+1}{r}\right)/4.
\end{eqnarray*}
 
Notice that we denote explicitly the dependence of the gamma-MRFs on the value of $\{c_r\}_r>0$, which act as regularization parameters that control the amount of abundance spatial smoothness enforced by each gamma-MRF (and which can differ among endmembers). Following an empirical Bayesian approach, the value of each $c_r$ will remain unspecified and will be adjusted automatically during the inference procedure using maximum marginal likelihood estimation (see \cite{Altmann2016} for details).

\subsubsection{Anomaly model}
As in \cite{Newstadt2014ssp,Altmann2015c}, the outliers are assumed to be spatially and spectrally sparse, i.e., for most of the pixels and spectral bands there are no outliers. To model outlier sparsity, we factor each outlier vector as
\begin{eqnarray}
\label{eq:outliers_decomp}
\bfr_{i,j} = \bfz_{i,j} \odot \bfx_{i,j},
\end{eqnarray}
where $\bfz_{i,j}=[z_{i,j,1},\ldots,z_{i,j,L}]\transp \in \left\lbrace 0,1 \right\rbrace^{\nbband}$ is a binary label vector, $\bfx_{i,j} \in \bbR^{\nbband}$ and $\odot$ denotes the Hadamard (term-wise) product. This decomposition allows one to decouple the location of the sparse components from their values. More precisely, 
$z_{i,j,n}=1$ if an outlier is present in the $\ell$th spectral band of the pixel $(i,j)$ with value equal to $r_{i,j,\ell}=x_{i,j,\ell}$. 

Assuming that the potential anomalies \emph{a priori} share the same statistical properties, we consider the following independent conjugate gamma priors
\begin{eqnarray}
\label{eq:prior_X}
x_{i,j,\ell}|\alpha, \nu \sim \mathcal{G} (x_{i,j,\ell};\alpha, \nu), \quad \forall i,j,\ell,
\end{eqnarray}
where $(\alpha,\nu)$ are arbitrarily fixed parameters. In general it is difficult to empirically estimate outlier hyperparameters due to the fact that in most applications, outliers are by definition rare events, ocurring in only a few pixels, and the observed waveforms are have unknown sparsity levels. Setting $(\alpha,\nu)$ so that $\textrm{E}\left[x_{i,j}|(\alpha,\nu) \right]=\alpha\nu$ is too high might lead to poor detection performance (outliers not detected), in particular in the presence of low amplitude outliers. Conversely, setting $(\alpha,\nu)$ so that $\textrm{E}\left[x_{i,j}|(\alpha,\nu) \right]=\alpha\nu$ is too small might lead to high probabilities of false alarm. However, in our experiments (See Section VI), we did not observe significant performance degradation when varying $(\alpha,\nu)$ over the range considered in Section VI.

For many applications, the locations of outliers are likely to be spectrally correlated (e.g., water absorption bands) and/or spatially correlated (weakly represented components, local nonlinear mixtures,\ldots). An effective way to take correlated outliers/nonlinear effects into account is to use a Markov random field (MRF) as a joint prior for the anomaly labels in $\bfZ= \left\lbrace\bfz_{i,j,\ell}\right\rbrace_{i,j,\ell}$. In this paper, we use the Ising model proposed in \cite{Altmann2015c} for robust linear unmixing of HSIs to define the prior model for $\bfZ$. MRFs have the property that the conditional distribution of a label $z_{i,j,\ell}$ given the other labels of the image equals the conditional distribution of this label vector given only its neighbors, i.e., $\textrm{P}(z_{i,j,\ell}|\bfZ_{\backslash z_{i,j,\ell}})= \textrm{P}(z_{i,j,\ell}|\bfZ_{\mathcal{V}_{i,j,\ell}})$, where $\mathcal{V}_{i,j,\ell}$ is the index set of the neighbors of $z_{i,j,\ell}$, 
$\bfZ_{\backslash z_{i,j,\ell}}$
denotes the matrix $\bfZ$
whose element $z_{i,j,\ell}$ has been removed and
$\bfZ_{\mathcal{V}_{i,j,\ell}}$ 
is the subset of $\bfZ$ composed of the elements whose indexes belong to $\mathcal{V}_{i,j,\ell}$. 
In this study, we consider that the spatial and spectral correlations can be different and thus consider two different neighborhoods. We decompose the neighborhood $\mathcal{V}_{i,j,\ell}$ as 
$\mathcal{V}_{i,j,\ell} = \mathcal{V}_{i,j,\ell}^L \cup \mathcal{V}_{i,j,\ell,n}^N$ where $\mathcal{V}_{i,j,\ell}^N$ (resp. $\mathcal{V}_{i,j,\ell}^L$) denotes the spatial (resp. spectral) neighborhood of $z_{i,j,\ell}$.
Specifically, we consider an Ising model that can be expressed as 
\begin{eqnarray}
\label{eq:Potts}
\textrm{P}(\bfZ | \bbeta') & = & \dfrac{1}{B(\bbeta')} \exp \left[\bbeta\transp \bphi(\bfZ) + \phi_{0}\left(\bfZ,\beta_0\right)\right]
\end{eqnarray}
where $\bbeta=[\beta_{N},\beta_{L}]\transp$, $\bbeta'=[\bbeta\transp,\beta_0]\transp$ and
\begin{eqnarray}
\left\{
    \begin{array}{lll}
        \phi_{L}\left(\bfZ\right) & = & \sum_{i,j,\ell} \sum_{z_{i,j,\ell'} \in \mathcal{V}_{i,j,\ell}^L} \delta(z_{i,j,\ell} - z_{i,j,\ell'}),\nonumber\\
\phi_{N}\left(\bfZ\right) & = & \sum_{i,j,\ell} \sum_{z_{i',j',\ell} \in \mathcal{V}_{i,j,\ell}^N} \delta(z_{i,j,\ell} - z_{i,j,\ell}),\nonumber\\
\bphi(\bfZ) & = & [\phi_{L}\left(\bfZ\right),\phi_{N}\left(\bfZ\right)]\transp,\nonumber\\
\phi_{0}\left(\bfZ,\beta_0\right) & = & \beta_0\sum_{i,j,\ell}  (1-z_{i,j,\ell}) +(1-\beta_0)\sum_{i,j,\ell}  z_{i,j,\ell},\nonumber

    \end{array}
\right.
\end{eqnarray}
and $\delta(\cdot)$ denotes the Kronecker delta function. Moreover, $\beta_{N}>0$ and $\beta_{L}>0$ are hyperparameters that control the spatial and spectral granularity of the MRF and $0\leq \beta_0 \leq 1$ is an additional parameter that models the probability of having outliers in the image. Specifically, the higher the value of $\beta_0$, the lower is the probability of outliers in the data. In a similar fashion to the gamma-MRFs parameters $\{c_r\}$ in \eqref{eq:prior1_r}, the Ising model hyperparameters will be adjusted  with a reduced computational cost via maximum marginal likelihood estimation. Different spectral and spatial neighbourhoods can be used in \eqref{eq:Potts}. In this paper, we consider a 4-neighbour 2D structure to account for the spatial correlation and a 2-neighbour 1D structure for the spectral dimension.  

\subsection{Joint Posterior distribution}
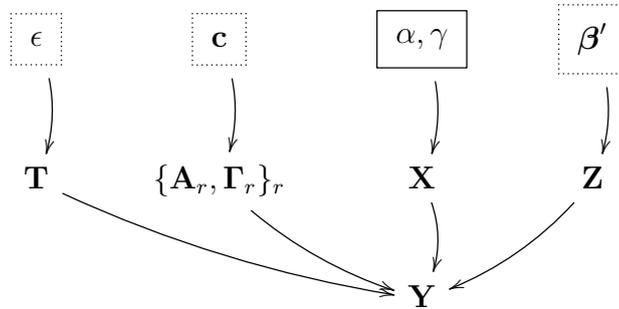
\begin{figure}[!ht]
\centerline{ \xymatrix{
 *+<0.05in>+[F.]+{\epsilon} \ar@/^/[d] & *+<0.05in>+[F.]+{\bfc} \ar@/^/[d] & *+<0.05in>+[F-]+{\alpha,\gamma} \ar@/^/[d]& *+<0.05in>+[F.]+{\bbeta'} \ar@/^/[d] \\
  \bfT \ar@/_/[rrd]    & \{\MATabond_r,\bGam_r\}_r \ar@/_/[rd] &   \bfX \ar@/^/[d]& \bfZ \ar@/^/[ld]   \\
   & & \MATpix & }
} \caption{Directed acyclic graph representing the proposed hierarchical Bayesian model. Fixed quantities appear in solid line boxes and quantities adjusted via maximum marginal likelihood estimation appear in dashed line boxes.} \label{fig:DAG}
\end{figure}

From the joint likelihood and prior model specified in Sections \ref{sec:lik} and \ref{sec:priors}, we can now derive the joint posterior distribution for $\bfT, \MATabond=\left\lbrace \MATabond_r \right\rbrace, \bGam=\left\lbrace \bGam_r \right\rbrace,\bfZ$, and $\bfX=\left\lbrace \bfx_{i,j}\right\rbrace$, given the observed waveforms $\MATpix$, and the value of the two sets of hyperparameters $\bPhi=(\alpha,\nu)$ and $\paramvect=(\bbeta', \epsilon, c_1,\ldots, c_{\nbmat})$. Note that the hyperparameters are organized into two groups: $\bPhi$ are the fixed variables and $\paramvect$ are those that will be adjusted. Using Bayes' theorem, and assuming prior independence between $\bfT$, $(\MATabond,\bGam)$, $\bfX$ and $\bfZ$, the joint posterior distribution associated with the proposed Bayesian model is given by\\
$f(\bfT, \MATabond, \bGam, \bfZ, \bfX,\btheta|\MATpix,\bPhi,\paramvect)$
\begin{eqnarray}
\label{eq:joint_post}
 \propto f(\MATpix|\bfT, \MATabond, \bfZ, \bfX)f(\MATabond,\bGam|\bfc)f(\bfT|\epsilon)f(\bfX|\alpha,\nu)f(\bfZ|\bbeta')
\end{eqnarray}
with $\bfc=[c_1,\ldots, c_{\nbmat}]\transp$. 
The directed acyclic graph (DAG) summarizing the structure of the proposed Bayesian model is depicted in Fig. \ref{fig:DAG}.

\section{Estimation strategy}
\label{sec:MCMC}
The posterior distribution \eqref{eq:joint_post} captures all of the information about the unknowns given the observed data and the priors on the unknown parameters. To perform joint depth estimation and spectral unmixing of the MSL data, we use the following four different Bayesian estimators: 1) the minimum mean square error estimator (MMSE) of the abundances, 2) the marginal maximum a posteriori (MMAP) estimator of the anomaly labels, 3) conditioned on the estimated outliers location, the MMSE estimator of the anomaly values (in a similar fashion to \cite{Altmann2014a}) and 4) conditioned on the estimated outliers and abundances, the MMAP estimator of the target ranges. Note that we use the MMAP estimators for the target ranges and labels, as this Bayesian estimator is particularly well adapted for estimation of discrete parameters.

In order to approximate these estimators, we propose a Markov chain Monte Carlo method to generate samples according to the joint posterior 
\begin{eqnarray}
\label{eq:joint_post_MMLE}
f(\bfT, \MATabond, \bGam, \bfZ, \bfX|\MATpix,\bPhi,\hat{\paramvect}), 
\end{eqnarray} 
where $\hat{\paramvect}$ denotes the maximum marginal likelihood estimator of the Ising and gamma-MRFs regularisation
hyperparameter vector $\paramvect$ given the observed data $\MATpix$, i.e.,
\begin{eqnarray}
\label{eq:MMLE}
\hat{\paramvect} = \underset{\paramvect \in \Theta}{\textrm{argmax}} f(\MATpix | \bPhi, \paramvect),
\end{eqnarray}
where $\Theta$ is the admissible set for $\paramvect$. This can be viewed as an empirical Bayes approach for specifying
$\paramvect$ where hyperparameters with unknown values are replaced by point
estimates computed from observed data (as opposed to being
fixed a priori or integrated out of the model by marginalisation) \cite{Pereyra2014ssp}.
This strategy has several important advantages for variables such as $\paramvect$ whose
conditional distributions are intractable. In particular, it has significantly lower computational cost compared to that of competing approaches, such as ones that marginalize the posterior \eqref{eq:joint_post_MMLE} over $\paramvect$ during the inference procedure \cite{Pereyra2013ip}.

To sample from the posterior \eqref{eq:joint_post_MMLE}, we use a Metropolis-within-Gibbs sampler that includes Hamiltonian Monte Carlo (HMC) updates \cite{Brooks2011}. This sampling procedure generates sequentially the unknown parameters from their estimated conditional distributions. The corresponding update steps are specified in the remainder of this section. 

\subsection{Sampling the depth parameters} 
Sampling  the  target  ranges  can  be  achieved by  sampling  sequentially  each  depth  from  its  conditional
distribution
\begin{eqnarray}
\label{eq:post_T0}
f(t_{i,j}|\Vpix{i,j},\blambda_{i,j}, \bfT_{\mathcal{V}(i,j)} \paramvect)
\end{eqnarray}
i.e.,  by drawing randomly from discrete (with finite support $\bbT$) distributions. In a similar fashion to \cite{Altmann2016}, we use a Gibbs sampler  implemented  using  a  colouring  scheme  such  that many depths can be updated in parallel (2 steps required when considering a 1-order neighborhood structure). Note that if the prior model in \eqref{eq:prior_t2} is used instead of \eqref{eq:prior_t1}, it can be seen for \eqref{eq:joint_post} that 
\begin{eqnarray}
\label{eq:post_T}
f(\bfT|\MATpix,\MATabond, \bGam, \bfZ, \bfX,\bPhi,\paramvect)=\prod_{i,j} f(t_{i,j}|\Vpix{i,j},\blambda_{i,j})
\end{eqnarray}
with $f(t_{i,j}|\Vpix{i,j},\blambda_{i,j})\propto f(\Vpix{i,j}|t_{i,j},\blambda_{i,j})/T'$. Thus, all the depth parameters can be updated independently and simultaneously.

\subsection{Sampling the anomaly labels}
From \eqref{eq:joint_post}, we obtain
\begin{eqnarray}
f(z_{i,j,\ell}=k|\MATpix,\MATabond,\bfX,\bfZ_{\backslash z_{i,j,\ell}},\bbeta') \propto \tilde{\pi}_{i,j,\ell}^{(k)},\quad \forall (i,j,\ell),
\end{eqnarray}
where $k \in \{0,1\}$,
\begin{eqnarray}
\label{eq:post_z}
\log\left(\tilde{\pi}_{i,j,\ell}^{(k)}\right) & = & \sum_{\nobin=1}^{\nbbin}\log\left(f(\pix{i,j,\ell}{\nobin}|\lambda_{i,j,\ell}^{(k)}, t_{i,j}) \right) \nonumber\\
& - & \bbeta\transp \bphi(\bfZ) - \phi_{0}\left(\bfZ,\beta_0\right),
\end{eqnarray}
and $\lambda_{i,j,\ell}^{(k)}=\Vmat{\ell,:}\Vabond{i,j} +k x_{i,j,\ell}$ with $\Vmat{\ell,:}$ the $\ell$th row of $\MATmat$.
Consequently, the label $z_{i,j,\ell}$  can be drawn from its conditional distribution by drawing randomly from $\{0,1\}$ with probabilities given by 
\begin{eqnarray}
\label{eq:post_Z}
f(z_{i,j,\ell}=k|\MATpix,\MATabond,\bfX,\bfZ_{\backslash z_{i,j,\ell}},\bbeta') = \dfrac{\tilde{\pi}_{n,\ell}^{(k)}}{\tilde{\pi}_{i,j,\ell}^{(0)}+\tilde{\pi}_{i,j,\ell}^{(1)}}.
\end{eqnarray}
In our experiments we incorporated a Gibbs sampler implemented using a colouring scheme such that
labels which are not direct neighbours can be updated in parallel. 
It is important to mention here that the main computation cost associated with the label updates arises from the computation of the sum over $\nbbin$ in \eqref{eq:post_z}, especially from large values of $\nbbin$ (e.g., $\nbbin=3000$ in the experiments presented in Section \ref{sec:simulations}). Fortunately, this sum can be written (up to an addition constant that does not depend on $z_{i,j,\ell}$) as 
\begin{eqnarray}
\tilde{y}_{i,j,\ell} \log\left(\Vmat{\ell,:}\Vabond{i,j} +k x_{i,j,\ell} \right) - \left(\Vmat{\ell,:}\Vabond{i,j} +k x_{i,j,\ell} \right)\tilde{g}_{i,j,\ell}, \nonumber
\end{eqnarray}
where $\tilde{y}_{i,j,\ell}=\sum_{\nobin=1}^{\nbbin}\pix{i,j,\ell}{\nobin}$ and $\tilde{g}_{i,j,\ell}=\sum_{\nobin=1}^{\nbbin}g_{i,j,\ell}(\nobin-t_{i,j})$ are quantities that only need to be computed once (before running the sampler) and stored in a look-up table, as they do not depend on the unknown model parameters ($\tilde{g}_{i,j,\ell}$ can be evaluated in advance for any value of $t_{i,j} \in \bbT$).

\subsection{Sampling the abundances}
It is easy to show that\\
$f(\MATabond|\MATpix,\bfT, \bGam, \bfZ, \bfX,\bPhi,\paramvect)$
\begin{eqnarray}
\label{eq:post_A}
=\prod_{i,j} f(\Vabond{i,j}|\Vpix{i,j},t_{i,j},\bGam,\bfX,\bfZ,\bfc),
\end{eqnarray}
i.e., the abundances of the $N_{\textrm{row}}N_{\textrm{col}}$ pixels can be updated independently. Here we update the elements of each vector $\Vabond{i,j}$ simultaneously using a constrained Hamiltonian Monte Carlo update \cite{Brooks2011,Altmann2014a} for several reasons. First, the intrinsic correlation between the spectral signatures in $\MATmat$ (especially when considering materials spectrally similar) imposes strong correlations between the elements of $\Vabond{i,j}$, which slows down the convergence of algorithms updating the abundances sequentially. Second, by accounting for the local curvature of $f(\Vabond{i,j}|\Vpix{i,j},t_{i,j},\bGam,\bfX,\bfZ,\bfc)$, such an approach usually yields better convergence and mixing properties than standard random walks. Note that it can be shown that when $c_r>1, \forall r$, the conditional distribution $f(\Vabond{i,j}|\Vpix{i,j},t_{i,j},\bGam,\bfX,\bfZ,\bfc)$ is strictly log-concave, which makes the use of gradient-based updating schemes particularly well adapted. These constraints on the regularization parameters $c_r$ can easily be incorporated within the estimation process by specifying $\Theta$ in \eqref{eq:MMLE} accordingly and projecting the updated parameters onto $\Theta$ at the end of the update step (this step is not detailed here but the interested reader is invited to consult \cite{Pereyra2014ssp,Altmann2015c,Altmann2016} for further details).

\subsection{Sampling the anomaly values $\bfX$}
As the spatial/spectral organization of the anomalies is only encoded via the Potts model \eqref{eq:Potts}, it can be shown that
$f(\bfX|\MATpix,\MATabond,\bfT, \bGam, \bfZ,\bPhi,\paramvect)$
\begin{eqnarray}
\label{eq:post_X}
=\prod_{i,j,\ell} f(x_{i,j,\ell}|\Vpix{i,j},t_{i,j},\Vabond{i,j},\alpha,\nu,z_{i,j,\ell}).
\end{eqnarray}
when $z_{i,j,\ell}=1$, the conditional distribution $f(x_{i,j,\ell}|\Vpix{i,j},t_{i,j},\Vabond{i,j},\alpha,\nu,z_{i,j,\ell})$ is a non-standard distribution (although log-concave when $\alpha\geq 1$, similarly to the conditional distribution of the abundances), while it reduces to \eqref{eq:prior_X} when $z_{i,j,\ell}$. Here, to avoid additional algorithmic complexity associated with rejection sampling and random walk update schemes, we use a standard Metropolis-Hastings update for which the proposal distribution is the prior distribution \eqref{eq:prior_X}. When $z_{i,j,\ell}=0$, which is satisfied for most of the elements of $\bfX$ when the anomalies are indeed sparse, this update allows the candidates to be automatically accepted. In practice, we observed that this updating approach leads to reasonable acceptance rates when $z_{i,j,\ell}=1$. 

\subsection{Sampling the auxiliary variables $\bGam$}

Simply, as $\bGam$ does not appear in \eqref{eq:joint_lik}, sampling the elements of $\bGam$ reduces to sampling from inverse-gamma distributions \eqref{eq:prior1_r3}. Note that due to the structure of the gamma-MRFs considered (the elements of each matrix $\bGam_r$ are not directly connected), these updates can be performed in a parallel manner.  

The resulting sampler is summarized in Algo. \ref{algo:algo1}. Its output is a set of $N_{\textrm{MC}}$ samples which are then used to approximate the Bayesian estimators of interest after sampler burn-in period which discards the $N_{\textrm{bi}}$ first samples. Here, the length of the burn-in period is determined from preliminary runs by visual inspection of the chains.  
Although there appear to be nested loops in Algo. \ref{algo:algo1}, the labels and target ranges can usually be grouped such that, within a group of labels or range parameters, the variables are conditionally independent and can thus be updated independently and in a parallel manner. With the neighbourhood structure used in \eqref{eq:TV_reg} and \eqref{eq:Potts}, only 2 sequential steps are required to update the binary labels, and the same applies for the range parameters, thus improving the convergence speed of the sampler and reducing computation. Moreover, the Hamiltonian Monte Carlo updates used to sample the abundances further improve the convergence speed of the sampler, when compared to standard Metropolis-Hastings updates.

\section{Relation to Poisson Factor Analysis}
\label{sec:PFA}
It is interesting to note from \eqref{eq:model0} that 
\begin{eqnarray}
\tilde{y}_{i,j,\ell}|\lambda_{i,j,\ell}, t_{i,j} \sim \mathcal{P}\left(\lambda_{i,j,\ell}\tilde{g}_{i,j,\ell} \right).
\end{eqnarray}
In other words, each integrated waveform (summed over the time bins) for each pixel and wavelength, follows a Poisson distribution whose mean depends only on the spectral parameters $\lambda_{i,j,\ell}=\Vmat{\ell,:}\Vabond{i,j}+r_{i,j,\ell}$, scaled by $\tilde{g}_{i,j,\ell}$ which contains only information about the range of the target. If we assume that $\tilde{g}_{i,j,\ell}=\sum_{\nobin=1}^{\nbbin}g_{i,j,\ell}(\nobin-t_{i,j})$ is constant for all possible values of $t_{i,j} \in \bbT$ (which occurs in practice when there are no truncations of the impulse responses at the boundaries of $(t_{min},t_{max})$), the problem can be reduced to a more standard spectral unmixing problem. This unmixing problem is over two spatial and one spectral dimension. If the spatial distortion of the instrument responses can be neglected, the observation model can be rewritten
\begin{eqnarray}
\label{eq:RPFA}
\widetilde{\MATpix} \sim \mathcal{P} \left(\widetilde{\MATmat} \widetilde{\MATabond} + \widetilde{\bfR} \right)
\end{eqnarray} 
where $\widetilde{\MATpix}$ is an $\nbband \times N_{\textrm{row}}N_{\textrm{col}}$ matrix with elements $\tilde{y}_{i,j,\ell}$, $\widetilde{\MATmat}$ corresponds to the endmember matrix whose columns have been scaled by $\tilde{g}_{i,j,\ell}$, $\widetilde{\MATabond}$ is the reshaped abundance matrix and $\widetilde{\bfR}$ is a sparse matrix representing the anomalies. Equation \eqref{eq:RPFA} is closely related to the Poisson Factor analysis (PFA) model \cite{Zhou2012} and to the robust non-negative matrix factorization model under a Poisson noise assumption \cite{Fevotte2015}. Unlike in factor analysis, we assume the endmember matrix $\MATmat$ is known. In addition to the temporal aspect of the problem considered, the proposed model extends the PFA model from \cite{Zhou2012} to account for spatial correlation and the presence of outliers/anomalies. In contrast to the simulation based method developed in \cite{Zhou2012} to sample the posterior and which introduces discrete latent variables to simplify the sampling procedure (most updates are standard Gibbs steps), we used accept-reject procedures based on constrained HMC moves. 
In contrast to \cite{Zhou2012}, where the columns of $\widetilde{\MATmat}$ were assumed to be uncorrelated, these columns can be correlated in the spectral unmixing problem considered in this paper.

\section{Experiments}
\label{sec:simulations}

\begin{figure}[h!]
\begin{minipage}[b]{.49\linewidth}
  \centering
\includegraphics[width=3.7cm]{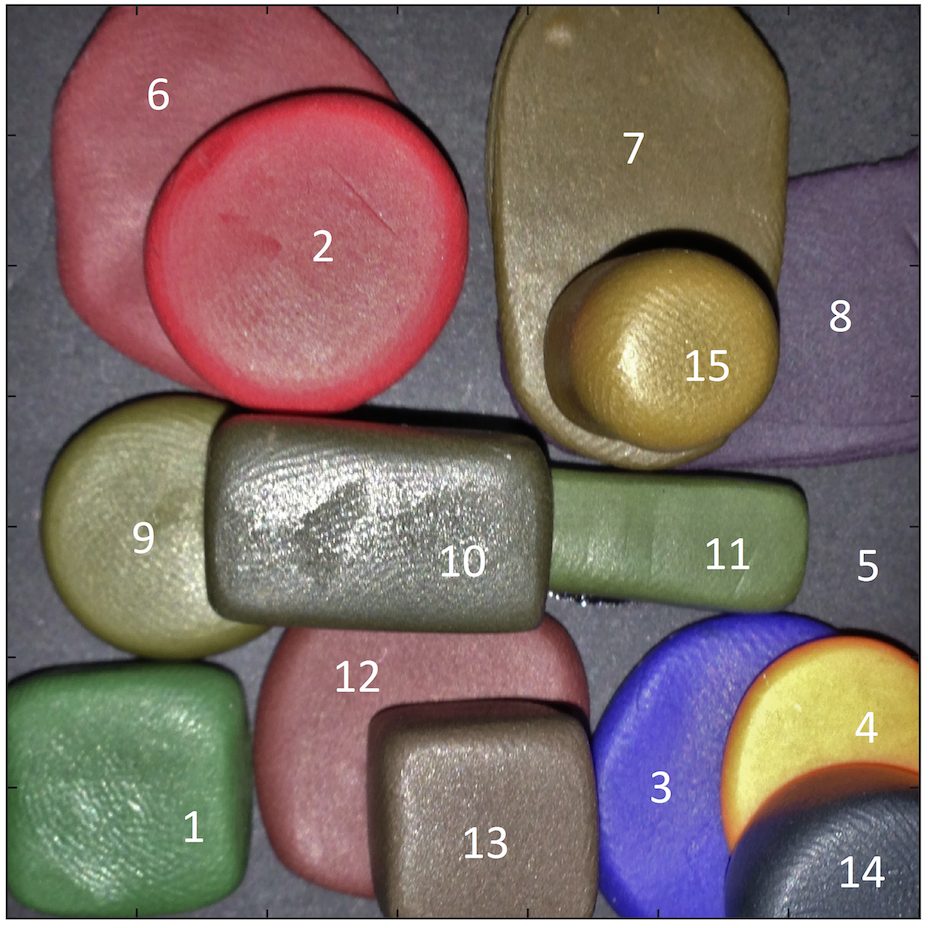}
  \centerline{(a) }\medskip
\end{minipage}
\hfill
\begin{minipage}[b]{0.49\linewidth}
  \centering
\includegraphics[width=4.23cm]{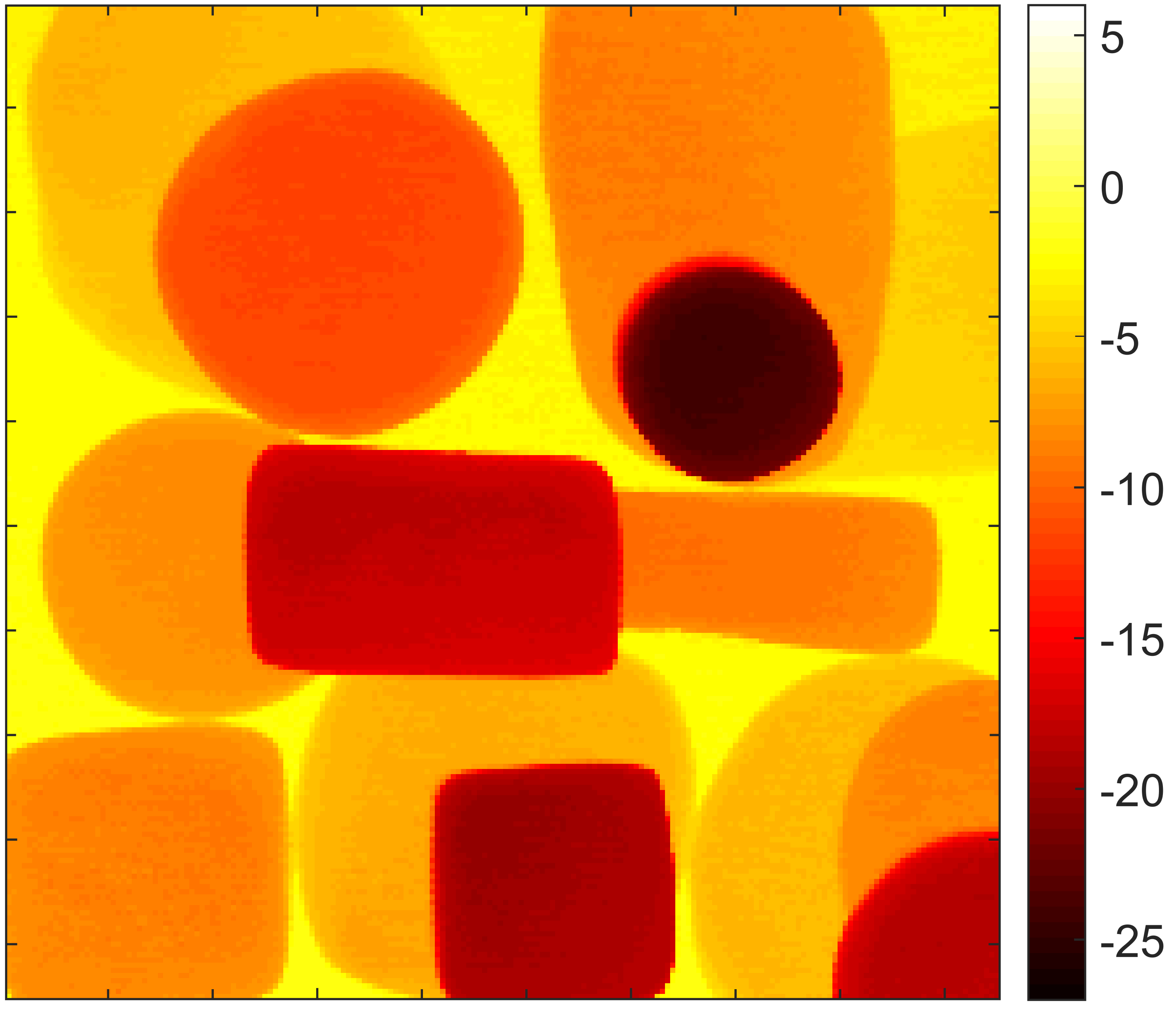}
  \centerline{(b)}\medskip
\end{minipage}
 \caption{(a): Standard RGB image of the first scene, composed of different coloured clays fixed on a dark-grey backboard. (b) Reference depth/range image in millimeter obtained using the algorithm proposed in \cite{ALTMANN_eusipco2016}.} \label{fig:blocks}
\end{figure}

\begin{figure}[h!]
  \centering
  \includegraphics[width=\columnwidth]{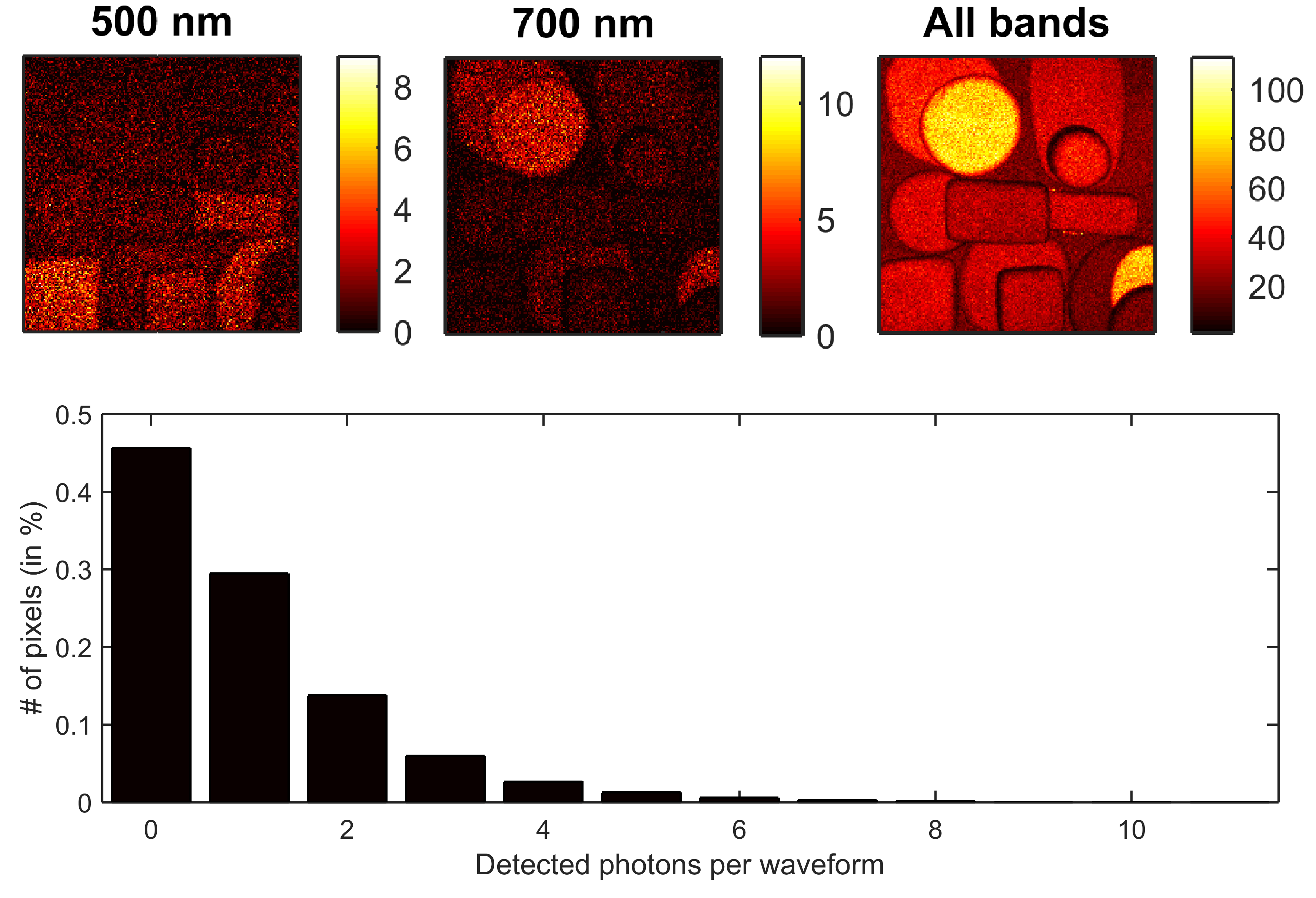}
  \caption{Distributions of the number of detected photons (integrated over the $T=3000$ bins) for the highest data sparsity level (on average $1$ photon per pixel for each band). Top: Spatial distribution of the detected photons at $500$nm (left), $700$nm (middle) and total photon counts after integration over the $L=33$ wavelengths (right). Bottom: Observed photon count distribution over all the pixels and bands. This figure shows that $45\%$ of the pixels do not contain any detection.}
  \label{fig:stat_data}
\end{figure}

\begin{figure}[h!]
  \centering
  \includegraphics[width=\columnwidth]{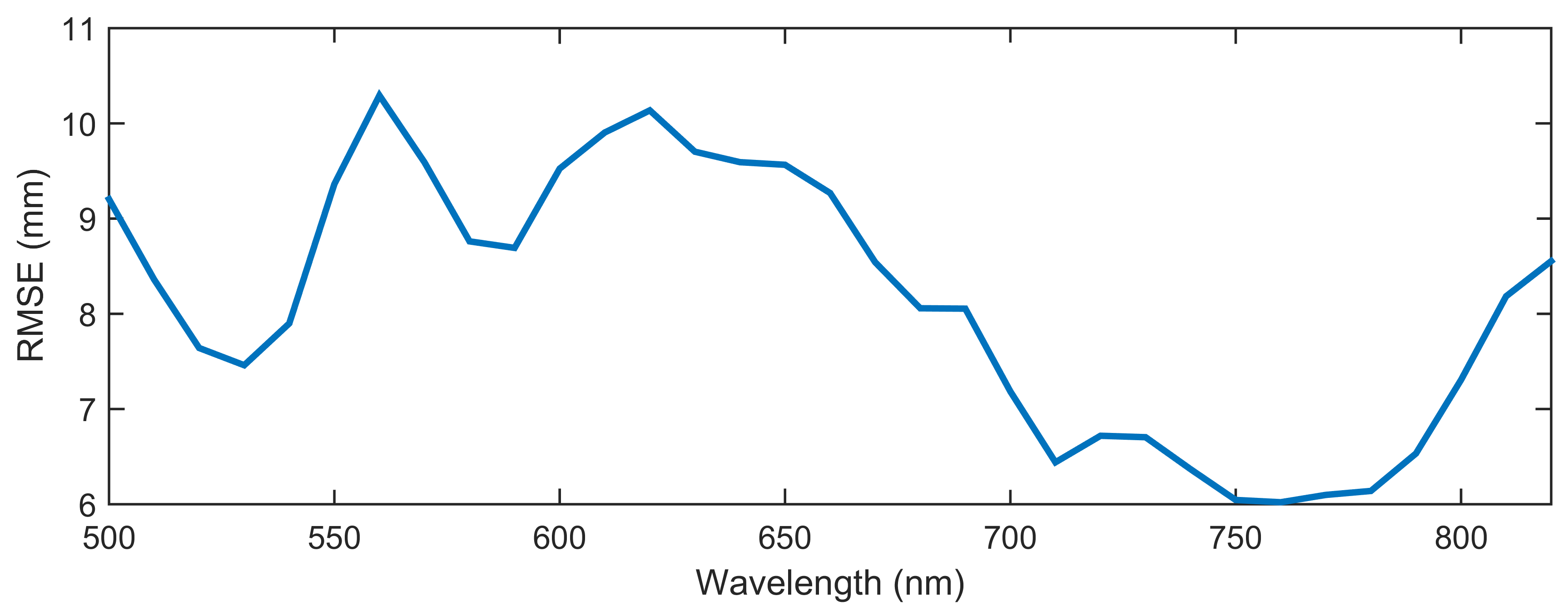}
  \caption{Depth RMSEs (in mm) obtained via maximum likelihood estimation when considering a single wavelength and using data constructed from $10$ photons per pixel (on average) for each wavelength. Target ranges of empty pixels are obtained via nearest-neighbour interpolation. These results illustrate how the depth estimation highly rely on the wavelength considered, especially when considering sparse data. Averaging the $L=33$ depth estimates generally reduces the RMSE and here, we obtained $\textrm{RMSE}=2.08$mm. }
  \label{fig:RMSE_10phot}
\end{figure}

We evaluate the performance of the proposed method applied to estimation of the depth and the spectral profiles of a $5 \times 5$cm  scene (see Fig. \ref{fig:blocks} (a)) composed of different objects made of opaque polymer clay and mounted on a dark-grey backboard at a distance of approximately $1.8$m from a time-of-flight scanning sensor. The sensor detects photons using a time-correlated single photon counting (TCSPC) technique.
The transceiver system and data acquisition hardware used for this work are similar to that described in \cite{McCarthy2009,Krichel2010,Wallace2010,McCarthy2013,Maccarone2015,ALTMANN2016ICASSP}, developed at Heriot-Watt University. The scanning system used was a monostatic transceiver that utilises a galvonometer mirror pair to define the field position and to direct the return signal to an individual silicon-based single-photon detector with picoseconds timing resolution. The measurements have been acquired indoors, under dark conditions to limit the influence of ambient illumination. The scene has been scanned using a regular spatial grid of $190 \times 190$ pixels and $L=33$ regularly spaced wavelengths ranging from $500$nm to $820$nm. Given the spatial structure of the scene considered in Fig. \ref{fig:blocks} (a) and the short distance between the detector and the target (constrained by indoor measurements), most of the pixels can be associated with a single material. In the sequel we use an $R=15$ endmember library/dictionary formed from the spectral response of each material and hence, for this experiment, almost all pixels in the image represent pure pixels without mixing. However, when the laser spot size on the target is broadened due to beam divergence then the assumption of a single material in each pixel is unlikely to hold. Such a situation would apply to longer range targets and close range targets whose material properties vary significantly over the spot size. Although classification methods such as \cite{Altmann2016SSP} can effectively be used to classify pixels in this particular scene, our aim is to demonstrate the ability of our method to discriminate the materials and estimate depth even in the very low photon count regime. The histograms consisted of $T=3000$ bins of $2$ps, which represents a depth resolution of $300\mu$m per bin. The optical power of the supercontinuum laser system was adjusted using data from preliminary runs, with a per-pixel acquisition time of $10$ms for each wavelength, in order to obtain accurate reference depth profiles (of approximately $1000$ photons per pixel at each spectral band on average). This corresponds to an overall acquisition time of approximately 3 hours and 20 minutes using $L=33$ sequential scans. The data format of timed events allows us to construct photon timing histograms associated with shorter acquisition times, as the system records the time of arrival of each detected photon (with respect to the previous synchronisation signal). Here, we evaluate our method for average photon counts of $1$, $3$ and $10$ photons per pixel for each wavelength. Examples of distributions of detected photon counts, for the shortest acquisition time, are depicted in Fig. \ref{fig:stat_data}. The top sub-figures in Fig. \ref{fig:stat_data} show that the number of detected photons varies depending both on the object and wavelength. Consequently, the depth estimation performance, which is highly dependent on the number of detected photons, can be significantly affected by the considered wavelength(s). This observation is confirmed by the results in Fig. \ref{fig:RMSE_10phot} which depicts the root mean squared error (RMSE) of the depth estimate, obtained using data constructed from $10$ photons per pixel (on average) for each wavelength and using a single wavelength to estimate the target ranges via maximum likelihood (ML) estimation. In this work the RMSE is defined by 
\begin{eqnarray}
\textrm{RMSE}=\sqrt{\dfrac{\sum_{i,j}(t_{i,j}-\hat{t}_{i,j})^2}{N_{\textrm{row}}N_{\textrm{col}}}},
\end{eqnarray}
where $\hat{t}_{i,j}$ is the estimated target range in the pixel $(i,j)$ and $t_{i,j}$ is the reference target range.  The reference depth profile shown in Fig. \ref{fig:blocks} (b) has been obtained using the algorithm proposed in \cite{ALTMANN_eusipco2016}. The algorithm provided for depth estimation and clustering the spectra in the MSL data. In a similar fashion to the method proposed in the paper, the algorithm considered in \cite{ALTMANN_eusipco2016} uses all the wavelengths to estimate the target ranges. However, it does not rely on a particular mixing model and is thus less sensitive to the presence of complex spectral mixtures. It was applied to the data with the longest acquisition time ($10$ms) to reduce estimation errors and the resulting reference range profile is in very good agreement with the structure of the scene in Fig. \ref{fig:blocks} (a) (the reference range being set to the range of the backboard).

\begin{figure}[h!]
  \centering
  \includegraphics[width=\columnwidth]{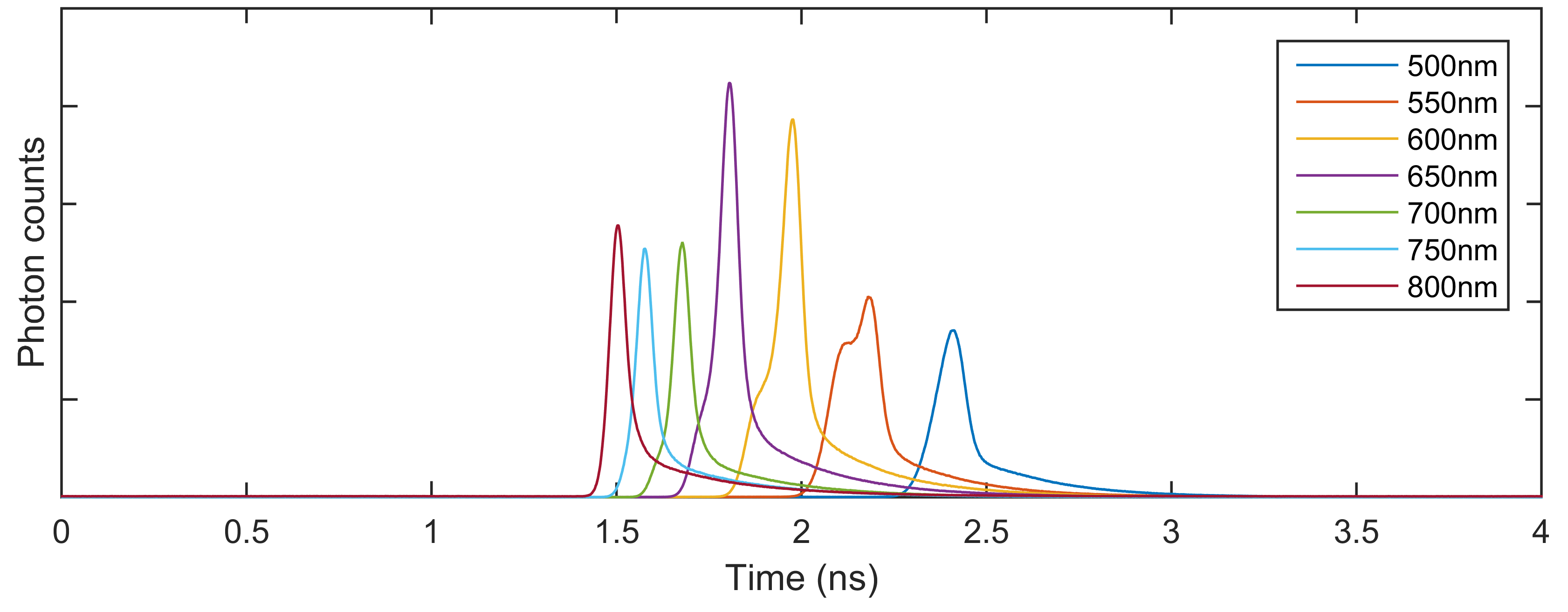}
  \caption{Examples of instrumental impulse responses measured with an acquisition of $100$s at different wavelengths ($500, 550, 600, 650, 700, 750$ and $800$nm).}
  \label{fig:IR}
\end{figure}

The instrument impulse responses $g_{i,j,\ell}(\cdot)$ (partly depicted in Fig. \ref{fig:IR} for an arbitrary central pixel of the field of view) were estimated from preliminary experiments by analyzing the distribution of photons reflected onto a Lambertian scatterer placed at a known distance with long acquisition time ($100$s). Fig. \ref{fig:IR} illustrates the fact that the response of the imaging system can change in amplitude and shape depending on the wavelength (full width at half maximum around $60$ps). This variability is due to the wavelength-dependent characteristics of the different elements in the imaging system, e.g., supercontinuum laser source, detector, lenses. Notice also the delays between the different peaks. These are mainly due to the wavelength-dependent response of the laser source and the different path lengths of the light in the imaging system. These delays can be compensated during the instrument calibration phase of the experiment and do not have a significant influence of the imaging performance. The spatial distortion of the responses $g_{i,j,\ell}(\cdot)$ due to potential limitations of the imaging system are taken into account during system calibration. If only a single wavelength was to be used to estimate the depth profile (e.g., as in Fig. \ref{fig:RMSE_10phot}), the variations of the peak shape could also make the choice of the most relevant wavelength difficult as the depth estimation accuracy mainly depends on the amplitude (reflectivity estimation) and width (depth estimation) of the peak.

\begin{figure}[h!]
  \centering
  \includegraphics[width=\columnwidth]{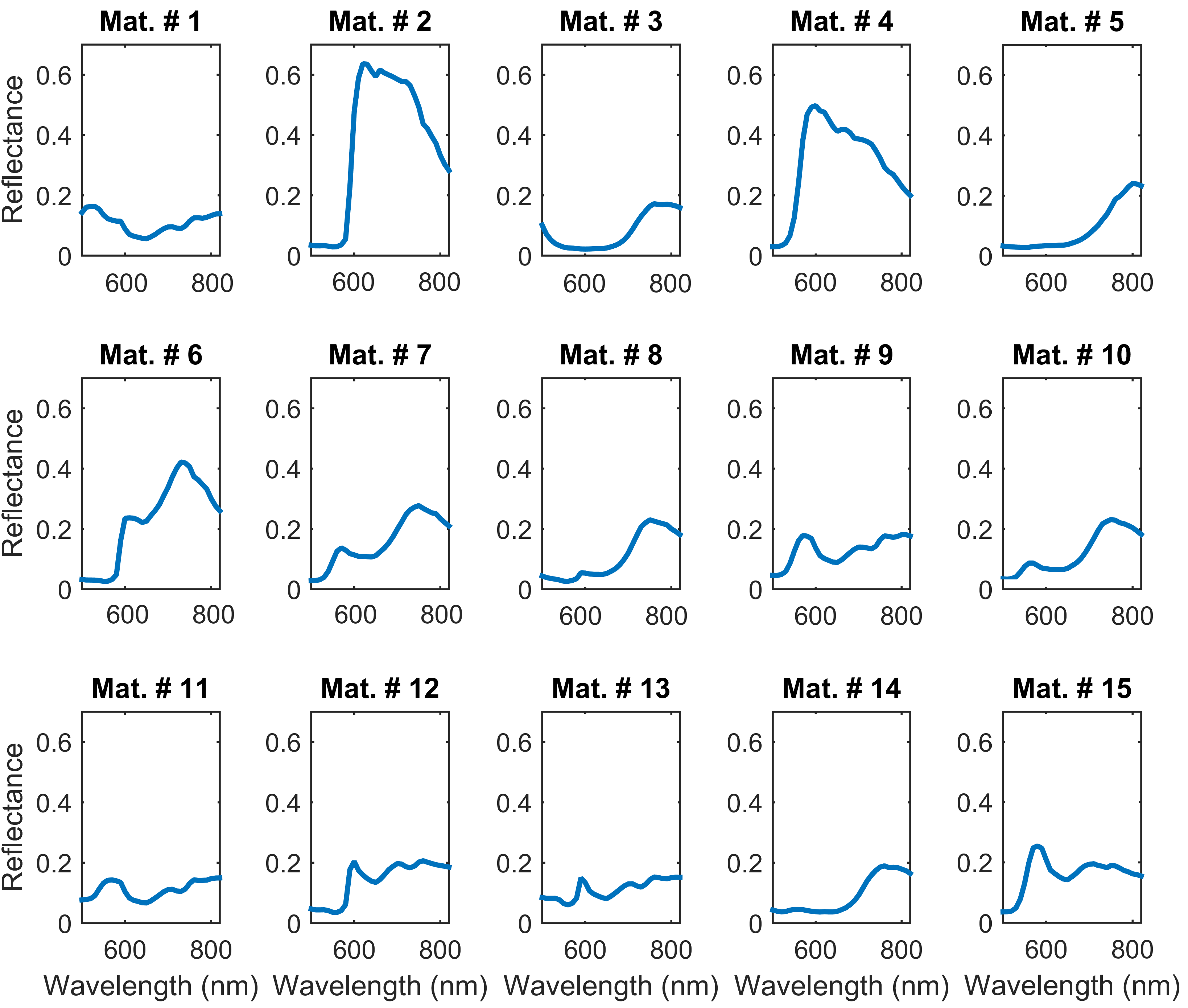}
  \caption{Spectral signatures of the backboard and the $14$ polymer clays used to create the objects in the scene shown in Fig. \ref{fig:blocks}.}
  \label{fig:endmembers}
\end{figure}

Fig. \ref{fig:endmembers} shows the spectral signatures of the backboard and the $14$ kinds of clay used to create the objects.
The $R=15$ endmembers were obtained using the algorithm proposed in \cite{ALTMANN_eusipco2016} for depth estimation and clustering on the
spectra in the MSL data. To ensure high accuracy of the endmembers they were extracted from much cleaner data, i.e., images of the sample acquired with much longer acquisition time resulting in  more than $1000$ photons per pixel and per band on average. To be more explicit, the endmembers were obtained by averaging the estimated spectral responses over sub-regions ($400$ pixels per endmember) identified by the method proposed in \cite{ALTMANN_eusipco2016}.

For all results presented here, the proposed Bayesian Poisson unmixing algorithm has been applied with $N_{\textrm{MC}}=5000$ sampler iterations, including $N_{\textrm{bi}}=2000$ burn-in iterations. These parameters were determined from preliminary runs by visual analysis of the chains and the variation of the results over independent runs. Moreover, we set $(\alpha,\nu)=(1,0.05)$ for all the simulation results presented in this Section. These parameters were selected based on the expectation that anomalies present low reflectivities. The results were not appreciably affected by small variation of ($\alpha, \nu$) about these values.

\begin{figure}[h!]
  \centering
  \includegraphics[width=0.8\columnwidth]{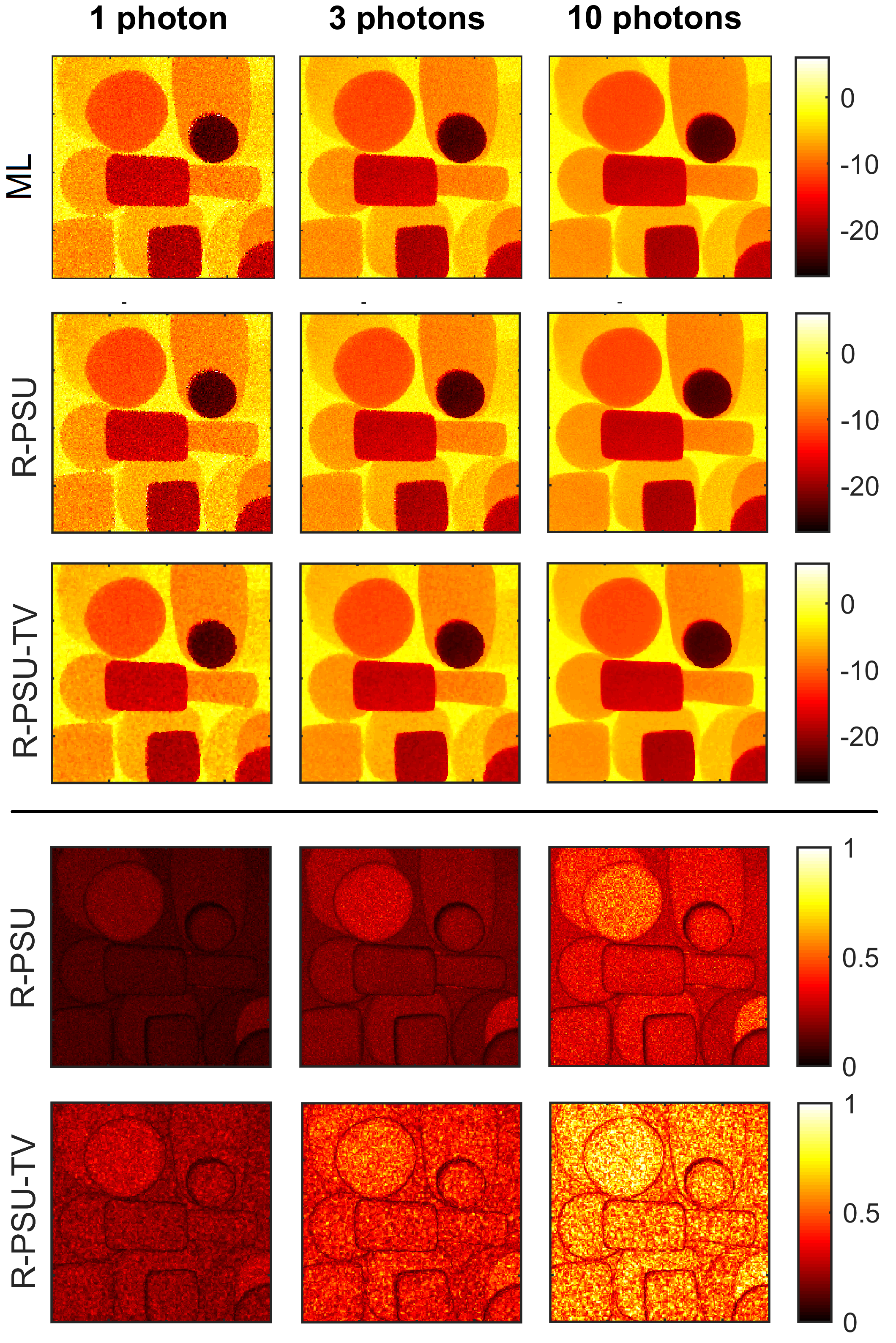}
  \caption{Top rows: Depth profiles estimated via pixel-wise ML estimation \cite{Altmann2016whispers} and the proposed method, with and without spatial regularization of the depth profile. Bottom rows: Confidence map, which is the marginal posterior probability, for each pixel, that the actual object range is within the estimated range bin. The higher this probability, the more reliable the estimated depth.}
 \label{fig:compare_depth}
\end{figure}

\begin{figure}[h!]
  \centering
  \includegraphics[width=\columnwidth]{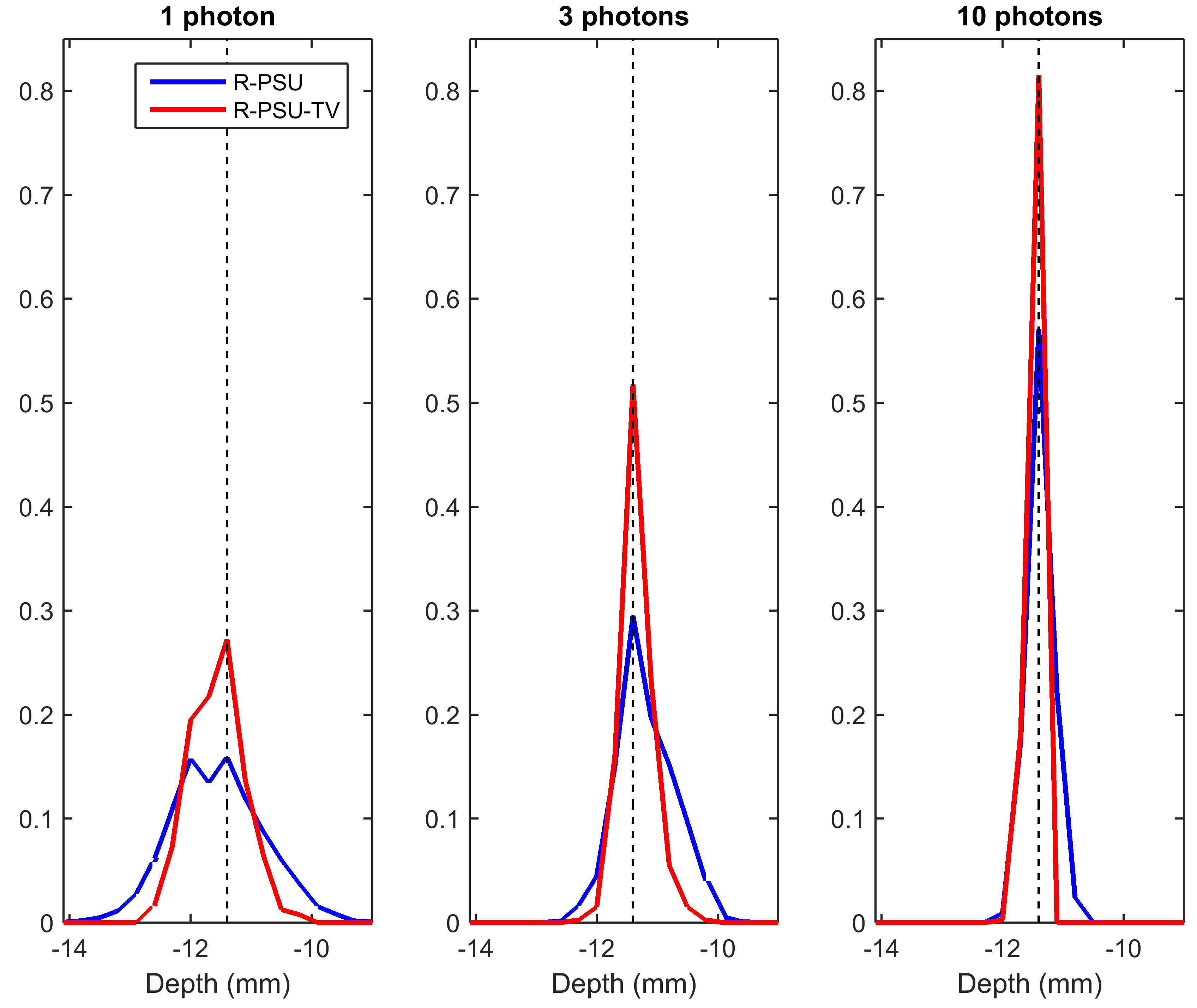}
  \caption{Marginal posterior distributions of the object range estimated using R-PSU (blue lines) and R-PSU-TV (red lines), for the pixel $(41,61)$ (central region of the object $\#2$). The black dashed lines represent the reference range estimated using the algorithm proposed in \cite{ALTMANN_eusipco2016}. Total-variation (R-PSU-TV) provides a higher posterior confidence on the estimated depth.}
 \label{fig:compare_depth_posterior}
\end{figure}

The depth profiles estimated by the proposed algorithm, referred to as R-PSU (for Robust Poisson Spectral Unmixing) are depicted in the second and third top rows of Fig. \ref{fig:compare_depth}. R-PSU-TV denotes the algorithm with the TV-based depth regularization (and R-PSU without TV-based regularization). For completeness, the top row of Fig. \ref{fig:compare_depth}, depicts the results obtained by the unmixing method proposed in \cite{Altmann2016whispers}, relying on the lineariarity of the mixtures. In contrast to the proposed method, the method in \cite{Altmann2016whispers} unmixes the pixels independently using weakly informative abundance and depth prior models. Consequently, this method is denoted by ML in Fig. \ref{fig:compare_depth}. As expected, the estimated depth profile becomes noisier as the average number of detected photons decreases. However, the TV-based depth prior model visually improves the estimated profiles, in particular for the highest data sparsity ($1$ photon on average per pixel and per band). The bottom rows of Fig. \ref{fig:compare_depth} depict the marginal posterior probabilities, for each pixel, of the actual object range to be within the estimated range bin (i.e., within a $0.3$mm interval centred around the estimated depth value). These images show that these probabilities increase with the number of detected photons, i.e., the joint posterior distribution becomes more concentrated around its modes and that the TV-based prior further concentrates this posterior distribution, thus reducing the \emph{a posteriori} uncertainty. This observation is also illustrated in Fig. \ref{fig:compare_depth_posterior}, which depicts the marginal posterior distributions of an arbitrary depth parameter (namely of the pixel $(41,61)$ chosen randomly), obtained using R-PSU and R-PSU-TV. Note that in the bottom rows of Fig. \ref{fig:compare_depth}, the lowest probabilities are generally found at the object boundaries, where significant range changes occur and where the number of detected photons decreases due to the local orientation of the objects. These two effects generally increase the estimation uncertainty. Moreover, the RMSEs gathered in Table \ref{tab:RMSEs} confirm the performance degradation as the number of detected photons decreases and that the TV-based prior model for the depth parameters mitigates this degradation (the RMSE remains below $1$mm, even for the highest data sparsity). These results also illustrate the benefits of the proposed Bayesian approach for depth estimation since the RMSE obtained with the highest data sparsity ($1$ photons per pixel and per band) and R-PSU-TV ($\textrm{RMSE}=0.92$mm) is much lower than the RMSE obtained by the joint ML depth estimates proposed in \cite{Altmann2016whispers}, which processes the pixels independently ($\textrm{RMSE}=3.66$mm).

\begin{figure*}
\begin{minipage}[b]{.49\linewidth}
  \centering
\includegraphics[width=8.5cm]{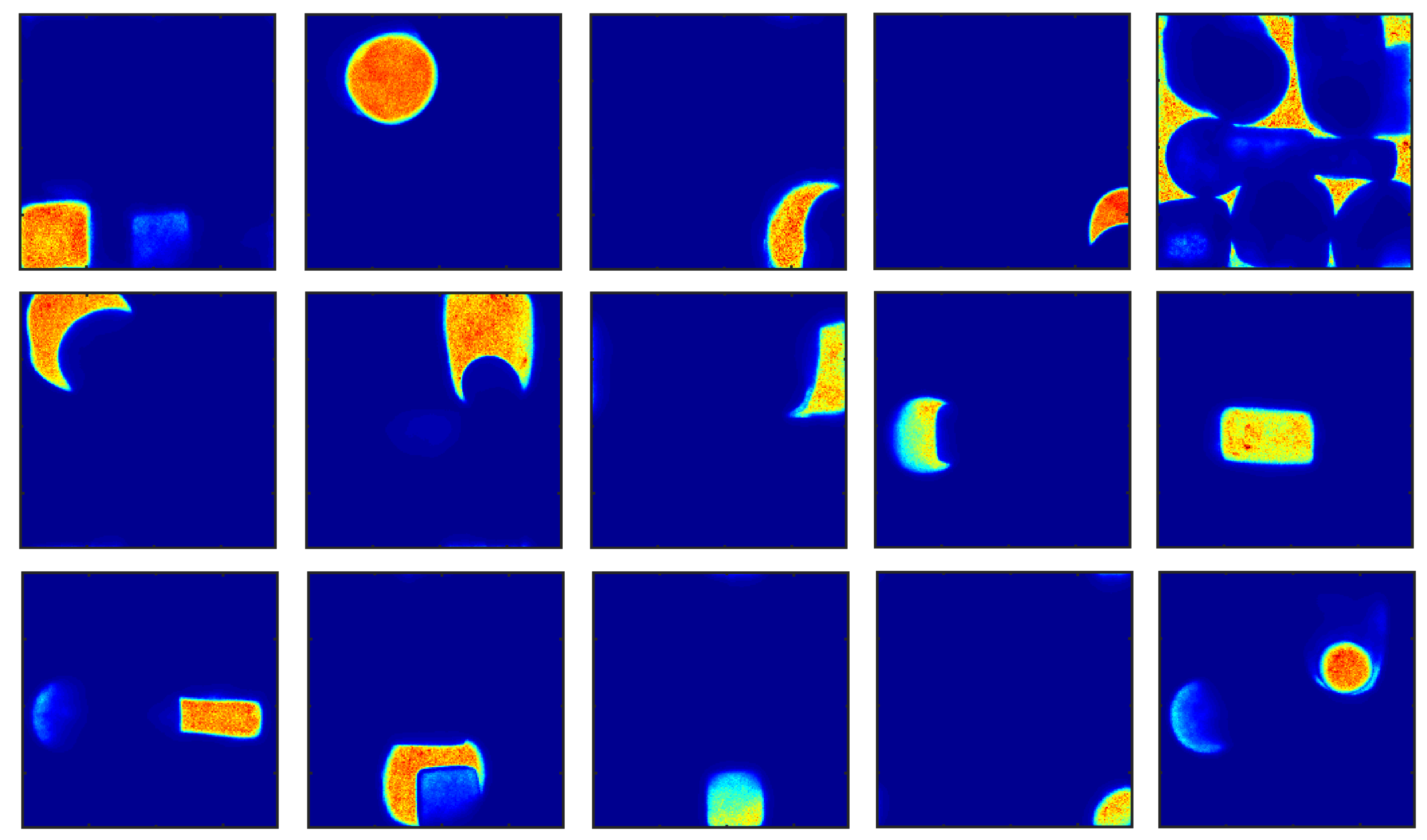}
  \centerline{(a) }\medskip
\end{minipage}
\hfill
\begin{minipage}[b]{0.49\linewidth}
  \centering
\includegraphics[width=8.5cm]{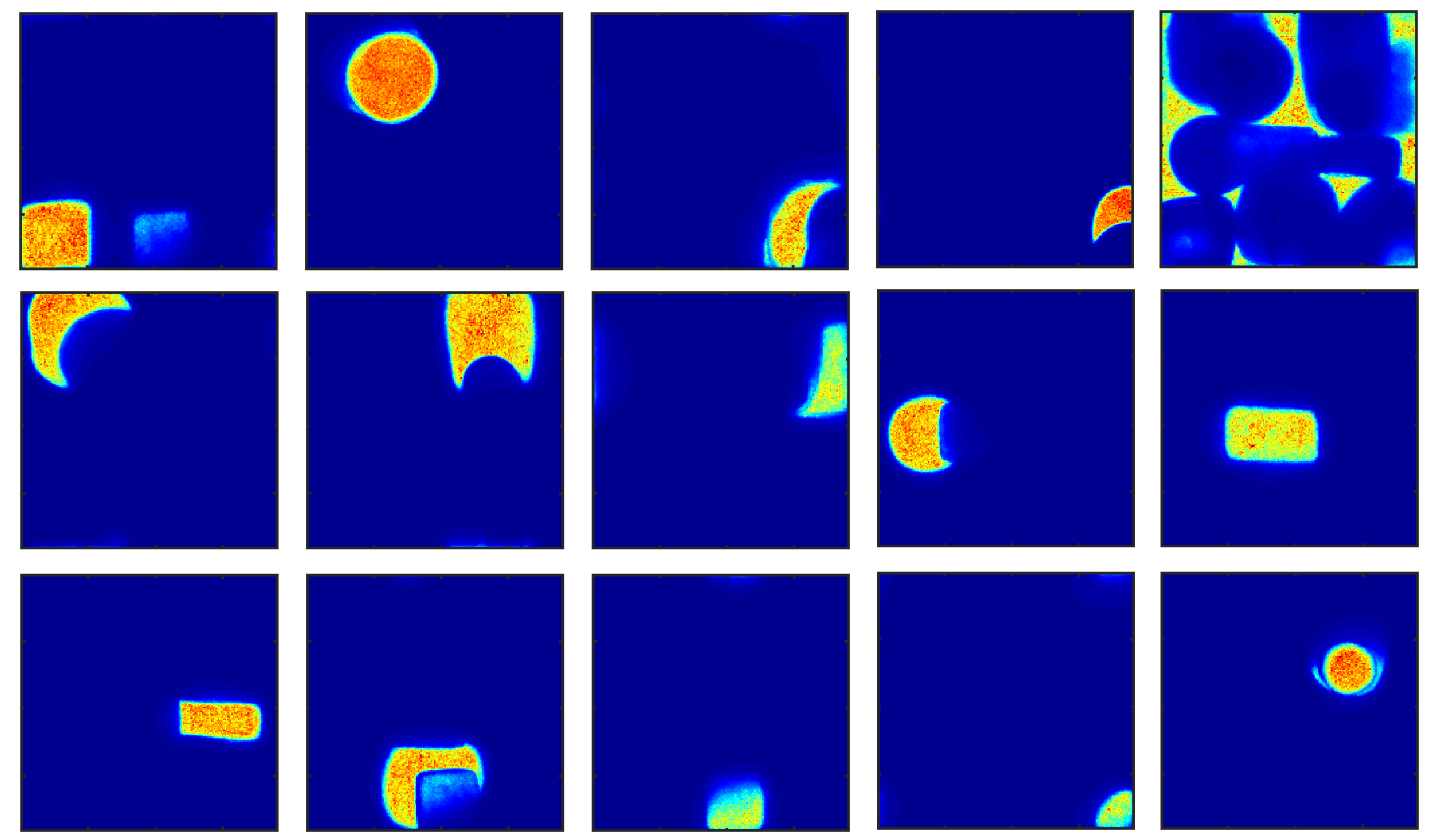}
  \centerline{(b)}\medskip
\end{minipage}

\begin{minipage}[b]{.49\linewidth}
  \centering
\includegraphics[width=8.5cm]{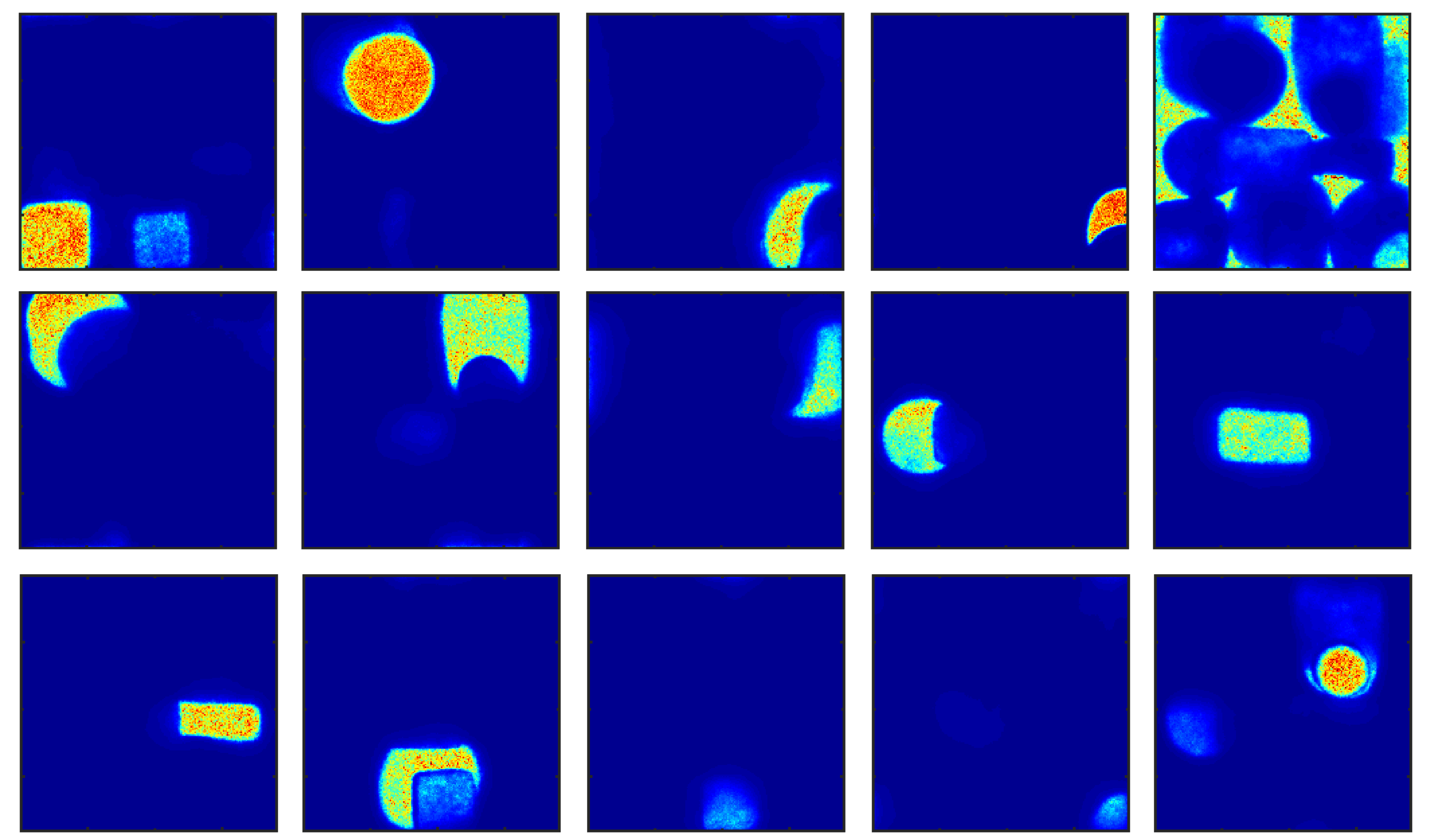}
  \centerline{(c) }\medskip
\end{minipage}
\hfill
\begin{minipage}[b]{0.49\linewidth}
  \centering
\includegraphics[width=8.5cm]{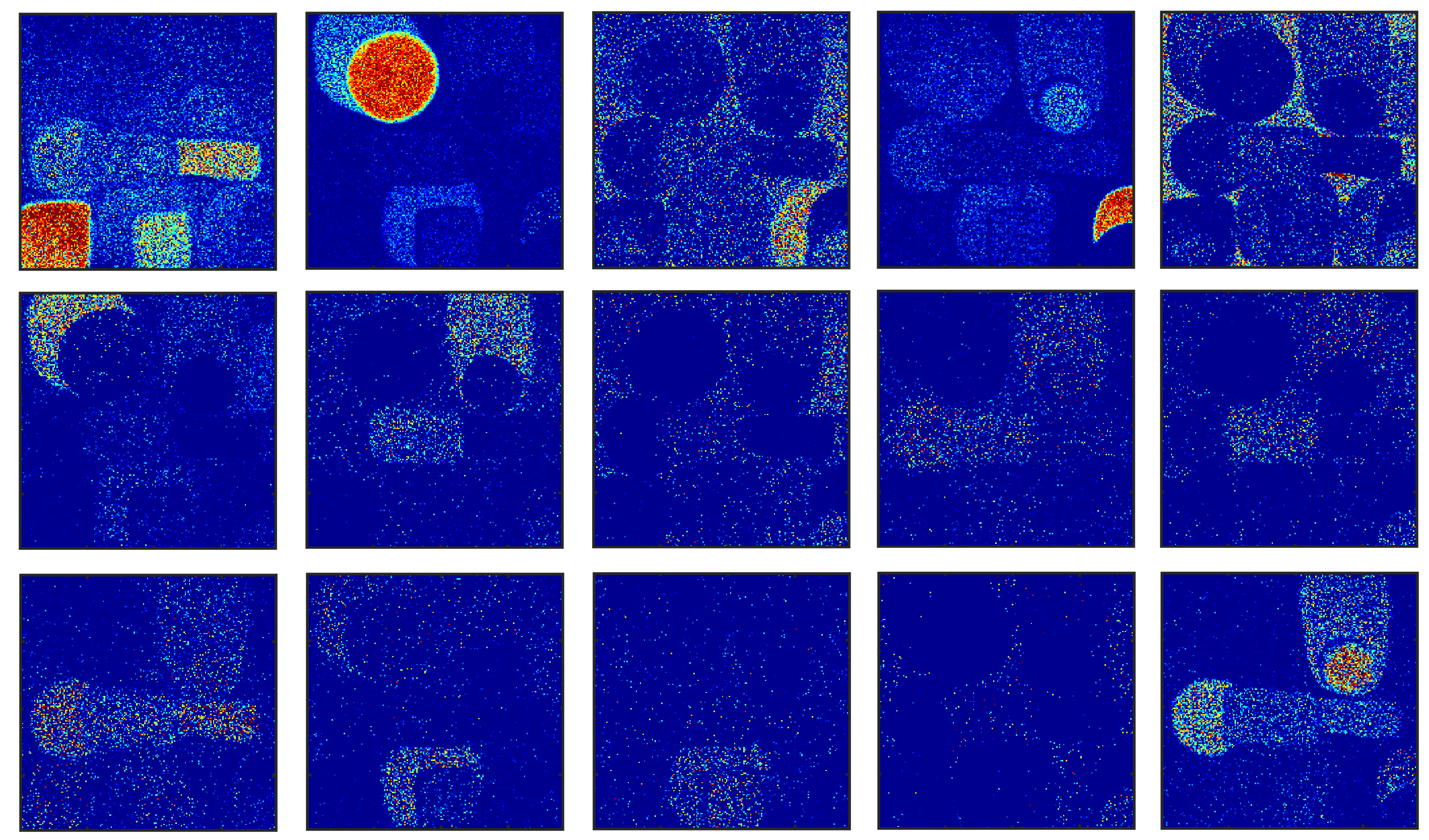}
  \centerline{(d)}\medskip
\end{minipage}
 \caption{(a)-(c): Abundance maps associated with the $R=15$ main materials composing the scene and estimated by the proposed R-PSU-TV method. These results are obtained from data constructed from $10$ (a), $3$ (b) and $1$ (c) photons per pixel (on average across the pixels) and for each spectral band. (d): Abundance maps estimated by the algorithm proposed in \cite{Altmann2016whispers} based on the shortest acquisition time ($1$ photon per pixel). All images have the same dynamic, i.e., between $0$ and $1.3$.} \label{fig:abund_R15}
\end{figure*}

Fig. \ref{fig:abund_R15} ((a)-(c)) compares the abundances estimated by the proposed method for the three photon sparsity levels considered ($10$, $3$ and $1$ photons per pixel for each band). The estimated abundances are generally in good agreement with the colour image in Fig. \ref{fig:blocks} (a) as it is possible to identify the regions where the different clays are present. However, some unmixing errors are also visible, even when considering the lowest photon sparsity in Fig. \ref{fig:abund_R15} (a). For instance, objects presenting different shades of green in Fig. \ref{fig:blocks} (a) might not be perfectly unmixed due to their spectral similarity and their intrinsic spectral variability mainly caused by local orientation changes. When considering higher data sparsity levels (see Fig. \ref{fig:abund_R15} ((b) and (c))), the spectral variability of each object becomes negligible compared to the actual Poisson noise level and the proposed method identifies more accurately the different homogeneous regions thanks to the use of the gamma-MRFs in \eqref{eq:GMRF}. To illustrate the benefits of the gamma-MRFs, we unmixed the data using the algorithm proposed in \cite{Altmann2016whispers} and which does not promote spatially correlated abundances. For illustration, Fig. \ref{fig:abund_R15} (d) presents the estimated abundance maps obtained using the highest data sparsity ($1$ photon per pixel on average). Without consideration of spatial correlation between abundance vectors, the estimated abundances maps become extremely noisy and it become impossible to clearly identify most of the homogeneous regions. In particular, the objects $\# 8$ and $\# 14$ are barely visible in Fig. \ref{fig:abund_R15} (d) while more distinguishable in Fig. \ref{fig:abund_R15} (c). Note that due to the spectral similarity between these objects and the board ($\# 5$), which is present in several regions on the scene, the objects $\# 8$ and $\# 14$ are also visible in the abundance map of the board in Fig. \ref{fig:abund_R15} (c). Although the unmixing results in Fig. \ref{fig:abund_R15} ((a)-(c)) present some errors, the estimates abundances maps are more accurate than those obtained without the abundance gamma-MRFs.

\begin{figure}[h!]
  \centering
  \includegraphics[width=\columnwidth]{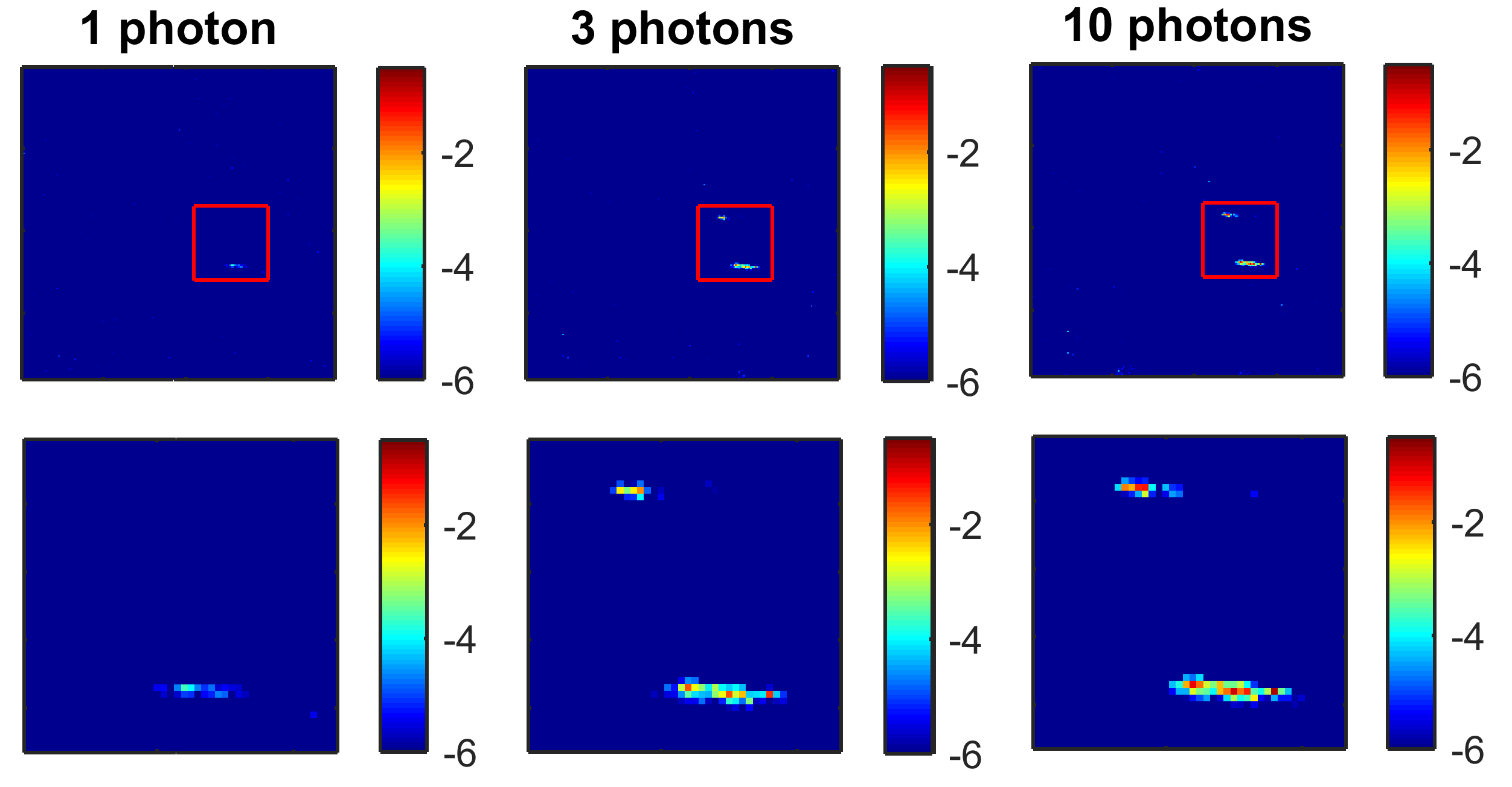}
 \caption{Anomaly maps ($\log(\norm{\bfr_{i,j}}^2/L)$) associated with the region of interest for averages of $1$, $3$ and $10$ detected photons per pixel and per band. The bottom row presents inset zooms of the squared red regions identified in the top row figures. The detected anomalies correspond to high reflectivity between $750$nm and $820$nm, and are due to the presence of residual glue used to fix the clay objects on the backboard.} \label{fig:blocks_anomaly}
\end{figure}

Fig. \ref{fig:blocks_anomaly} shows the estimated anomaly maps obtained using the $R=15$ endmembers of Fig. \ref{fig:endmembers} for the three acquisition times. This figure illustrates significant local differences from the classical LMM, in particular for the longest exposures. In the central region of the scene, below and above the object $\# 11$, the two red strips correspond to high reflectivity between $750$nm and $820$nm, and are due to the presence of residual glue used to fix the clay objects on the backboard. When the exposure decreases, the quality of the data decreases due to the statistical properties of the Poisson noise and it becomes more challenging to detect subtle spectral variations. In such a case, the proposed method, which promotes sparse deviations from the LMM, is no longer able to detect weak anomaly levels and only detects one glue strip (left sub-plots). It can be seen from Fig. \ref{fig:blocks_anomaly} that deviations from the standard LMM are clustered spatially and only affect a reduced number of pixels. Thus, the abundance maps estimated while considering the LMM (i.e., by enforcing $\norm{\bfr_{i,j}}^2=0$) are very similar, for most pixels, to those depicted in Fig. \ref{fig:abund_R15} and are not presented here for brevity.

\section{Conclusion}
\label{sec:conclusion}
We have proposed a new Bayesian model and a joint depth estimation and robust spectral unmixing algorithm for 3D scene analysis from MSL data.
Assuming the ambient illumination can be neglected, the spectra of the scene surfaces visible by the imaging system were decomposed into linear mixtures of known endmembers, potentially corrupted by sparse deviations/anomalies. Adopting a Bayesian approach, prior distributions were assigned to the unknown model parameters; in particular, a 3D Ising model was used to model the spatial organization of the anomalies and gamma Markov random fields were consider to promote spatially smooth abundances.
Including ambient illumination and dark count levels in the observation model (as in \cite{Kirmani2014,Shin2014,ALTMANN2016ICASSP,Altmann2016}) is the obvious next step from a more general application (especially for long-range imaging applications) of the proposed method. 

In this work, the experiments were performed indoor and the surface visible in each pixel was small ($\approx 0.01$mm$^2$) compared to the size of the objects. Applying the proposed method to large-standoff outdoor applications is under investigation. In such cases, the divergence of the laser sources can lead to changes in the shape of the returned pulses due to the local orientation of the observed objects with respect to the imaging system. Note that such distortion might also occurs when there are semi-transparent objects. The divergence of the laser sources, which increases with the target range, results in reduced spatial resolution leading to increased mixing within a single pixel.

In analogy to passive hyperspectral image analysis, endmember and mixture characterization  will become more challenging for large target ranges. Unsupervised or blind spectral unmixing approaches will need to be developed that can deal with unknown endmember spectral signatures. As explained in the introduction of this paper, our proposed anomaly model identifies ``anomalous''  pixels for which the linear mixing model  is a poor fit. Such pixels identified by our model can be subsequently processed by more sophisticated, and yet to be developed, non-linear mixing models. The development of such models is a worthy topic for future study.

Finally,  for remote sensing applications, it will be crucial to account for the presence of distributed (multi-layered or semi-transparent) targets, which would require more complex models for the multiple returns in the MSL data. This could potentially lead to full 3D abundance profile estimation.

\bibliographystyle{IEEEtran}
\bibliography{biblio}

\clearpage

\begin{table}[h!]
\renewcommand{\arraystretch}{1.2}
\begin{footnotesize}
\begin{center}
\begin{tabular}{|c|c|c|c|}
\cline{2-4}
\multicolumn{1}{c|}{} &  \multicolumn{3}{|c|}{Average no. of photons per pixel} \\
\cline{2-4}
\multicolumn{1}{c|}{}  & $1$ & $3$ & $10$ \\
\hline
ML & $3.66$ & $1.09$ & $0.65$\\
\hline
R-PSU & $1.91$ & $1.05$ & $0.65$\\
\hline
R-PSU-TV & $0.92$ & $0.64$ & $0.50$\\
\hline
\end{tabular}
\end{center}
\end{footnotesize}
\caption{Depth RMSEs (in mm) obtained via pixel-wise ML estimation \cite{Altmann2016whispers} and by the proposed method, with and without spatial regularization of the depth profiles.\label{tab:RMSEs}}
\vspace{-0.3cm}
\end{table}

\clearpage

\begin{algogo}{RSU-MSL algorithm}
     \label{algo:algo1}
     \begin{algorithmic}[1]
        \STATE \underline{Fixed input parameters:} Endmember matrix $\MATmat$, $(\alpha,\nu)$ number of burn-in iterations $N_{\textrm{bi}}$, total number of iterations $N_{\textrm{MC}}$
				\STATE \underline{Initialization ($u=0$)}
        \begin{itemize}
        \item Set $\bfT^{(0)}, \MATabond^{(0)},\bGam^{(0)},\bfX^{(0)},\bfZ^{(0)},\paramvect^{(0)}$
        \end{itemize}
        \STATE \underline{Iterations ($1 \leq u \leq N_{\textrm{MC}}$)}
        \STATE Sample $\MATabond^{(u)}$ from \eqref{eq:post_A}
        \STATE Sample $\bGam^{(u)}$ from \eqref{eq:prior1_r3}
				\FOR{$i=1:N_{\textrm{row}}$}
				\FOR{$j=1:N_{\textrm{col}}$}
				\STATE Sample $t_{i,j}^{(u)}$ from \eqref{eq:post_T0} or \eqref{eq:post_T}
				\FOR{$\ell=1:\nbband$}
				\STATE Sample $z_{i,j,\ell}^{(u)}$ from \eqref{eq:post_Z} 
				\ENDFOR
				\ENDFOR
				\ENDFOR
        \STATE Sample $\bfX^{(u)}$ from \eqref{eq:post_X}
				\IF{$u<N_{\textrm{bi}}$} 
				\STATE Update $\paramvect^{(u)}$ using \cite{Pereyra2014ssp} 
				\ELSE
				\STATE Set $\paramvect^{(u)}=\paramvect^{(u-1)}$
				\ENDIF
        \STATE Set $u = u+1$.
\end{algorithmic}
\end{algogo}

\end{document}

%% file: notations_yoann.tex
\input com_notations.tex

\newcommand{\Vpix}[1]{\mathbf{y}_{#1}}

\newcommand{\MATpix}{\mathbf{Y}}
\newcommand{\pix}[2]{y_{#1,#2}}

\newcommand{\nbband}{L}

\newcommand{\nbbin}{T}
\newcommand{\nobin}{t}

\newcommand{\nbmat}{R}
\newcommand{\nomat}{r}

\newcommand{\MATmat}{{\mathbf M}}
\newcommand{\Vmat}[1]{{\mathbf m}_{#1}}

\newcommand{\MATabond}{{\bold A}}

\newcommand{\abond}[2]{{a}_{#1,#2}}
\newcommand{\Vabond}[1]{{\boldsymbol{a}}_{#1}}

\newcommand{\paramvect}{\boldsymbol{\theta}}

\newcommand{\transp}{^T}

\newcommand{\norm}[1]{\left\|#1\right\|}

\newenvironment{algogo}[1]{
\smallskip
\noindent \hrule\vspace{0.2\baselineskip} \hrule
\begin{small}
\refstepcounter{algo} \center{\bf \textsc{Algorithm \thealgo}}
\\{\center{\bf #1}}
\smallskip
\flushleft
 } {
\end{small}
\smallskip
\hrule\vspace{0.2\baselineskip} \hrule
\smallskip
}

\newcounter{algo}
\renewcommand{\thealgo}{\arabic{algo}}

%% file: com_notations.tex
\def\bfc{{\mathbf{c}}}

\def\bfr{{\mathbf{r}}}

\def\bfx{{\mathbf{x}}}

\def\bfz{{\mathbf{z}}}

\def\bfR{{\mathbf{R}}}

\def\bfT{{\mathbf{T}}}

\def\bfX{{\mathbf{X}}}

\def\bfZ{{\mathbf{Z}}}

\def\bbR{{\mathbb{R}}}

\def\bbT{{\mathbb{T}}}
